\documentclass{emulateapj} 
\usepackage{amssymb, amsmath, apjfonts, natbib, color} 
\usepackage{rotating, graphics}

\newdimen\digitwidth 
\setbox1=\hbox{0} 
\digitwidth=\wd1 
\catcode`"=\active 
\def"{\kern\digitwidth} 
\slugcomment{Version 4.1; Accepted for publication in ApJ} 

\begin{document} 
\title{CIV EMISSION LINE PROPERTIES AND UNCERTAINTIES IN BLACK HOLE MASS ESTIMATES of $\lowercase{z}\sim3.5$ QUASARS} 
\author{Wenwen Zuo\altaffilmark{1}, Xue-Bing Wu\altaffilmark{2, 3}, Xiaohui Fan\altaffilmark{4}, Richard Green\altaffilmark{4}, Weiming Yi\altaffilmark{5, 6}, Andreas Schulze\altaffilmark{ 7}, Ran Wang\altaffilmark{3}, Fuyan Bian\altaffilmark{8}} 
\altaffiltext{1}{Shanghai Astronomical Observatory, Chinese Academy of Sciences, 80 Nandan Road, Shanghai 200030, China; wenwenzuo@shao.ac.cn} 
\altaffiltext{2}{Department of Astronomy, School of Physics, Peking University, Beijing 100871, China} 
\altaffiltext{3}{Kavli Institute for Astronomy and Astrophysics, Peking University, Beijing 100871, China}
\altaffiltext{4}{Steward Observatory, The University of Arizona, Tucson, AZ 85721, USA} 
\altaffiltext{5}{Yunnan Observatories, Kunming, 650216, China} 
\altaffiltext{6}{Key Laboratory for the Structure and Evolution of Celestial Objects, Chinese Academy of Sciences, Kunming 650216, China} 
\altaffiltext{7}{National Astronomical Observatory of Japan, Mitaka, Tokyo 181-8588, Japan} 
\altaffiltext{8}{European Southern Observatory, Alonso de Córdova 3107, Casilla 19001, Vitacura, Santiago 19, Chile}

\begin{abstract}
Using a high luminosity ($L_{\rm bol} \sim 10^{47.5} - 10^{48.3}$ erg s$^{-1}$), high redshift ($3.2 < z < 3.8$) quasar sample of 19 quasars with optical and near-infrared spectroscopy, we investigate the reliability of the CIV-based black hole mass estimates ($M_{\rm BH}$). The median logarithm of the CIV- and H$\beta$-based $M_{\rm BH}$ ratios is 0.110 dex with the scatter of 0.647 dex. The CIV-to-H$\beta$ BH mass differences are significantly correlated with the CIV FWHMs, blueshifts and asymmetries. Corrections of the CIV FWHM using the blueshift and asymmetry reduce the scatter of the mass differences by $\sim$ 0.04-0.2 dex. Quasars in our sample accrete at the Eddington ratio $R_{\rm EDD}>0.3$ and cover a considerable range of blueshifts, with 18/19 of the quasars showing CIV blueshifts (with the median value of 1126 km s$^{-1}$) and 14/19 of the quasars showing CIV blueshifts larger than 500 km s$^{-1}$. It suggests that not all quasars with high Eddington ratios show large blueshifts. The Baldwin effect between the CIV rest-frame equivalent width (REW) and the continuum luminosity at 1350 \AA\  is not seen, likely due to the limited luminosity range of our sample. We find a lack of flux in the red wing of the composite spectrum with larger CIV blueshift, and detect a higher ratio of [OIII] quasars with REW$_{\rm [OIII]}>5$ \AA\ in the subsample with lower CIV blueshift. It is more likely that they are caused by the combination of the Eddington ratio and the orientation effect.
\end{abstract}


\section{INTRODUCTION}

It is commonly accepted that black holes (BHs) reside in the center of massive galaxies, and the black hole mass ($M_{\rm BH}$) tightly correlates with properties of the host galaxy, i.e., the mass of the host galaxy spheroidal component and its velocity dispersion \citep{Ferrarese00, Gebhardt00, Tremaine02, Gultekin09, Graham11, Kormendy13}. These scaling relations strongly suggest that BH growth is coupled with galaxy mass assembly history \citep{Onken04, Nelson04, Kormendy13}. 
To understand the growth history of BHs and the connection between the BH and the host galaxy, it is important to obtain reliable BH mass estimates.

Assuming the line-emitting clouds in the broad-line region (BLR) are virialized under the 
gravity of the central BH, the BH mass can be estimated with the BLR size and the virial 
velocity of the BLR clouds. The Full Width at Half Maximum (FWHM) or the dispersion of the 
broad emission line is commonly used to represent the virial velocity. Mainly for low redshift 
Active Galactic Nuclei (AGN), the reverberation mapping (RM) technique has been applied; 
the time lag of the variations between the broad emission line and the continuum luminosity 
is used to trace the typical BLR size \citep{Peterson93, Peterson04}. 

However, the RM technique requires long term observational campaigns, and is especially challenging 
at higher redshift \citep{Kaspi_etal_00, Kaspi_etal_07, Peterson04, Bentz_etal_09, Denney_etal_10, 
Du_etal_14, Barth_etal_15, Shen_etal_15, Grier_etal_17, Grier_etal_19}. A tight correlation between 
the BLR size ($R$) and the quasar continuum luminosity in optical bands has been revealed from the 
RM campaigns \citep{Kaspi_etal_00, Kaspi_etal_05, Bentz_etal_13}. This relation provides an alternative inexpensive 
way to estimate the BLR size through single-epoch (SE) spectroscopy, further leading to the so-called 
SE virial BH mass estimates.

Under the virial assumption, BH masses for a large sample of AGNs can be calculated with the product of 
the BLR size and the virial velocity based on the SE spectroscopy via $M_{\rm BH} \propto R \times v^2$, 
with coefficients fairly well calibrated from ~40 $z<0.7$ AGNs with H$\beta$-based RM $M_{\rm BH}$ 
\citep{Kaspi_etal_00, Peterson04}. Typically, in the SE method the H$\beta$ broad emission line width and 
the continuum luminosity at 5100 \AA\ are used \citep[e.g.,][]{Greene05, Vestergaard_Peterson06, McGill08, Shen11, Shen_Liu12, Shen13, Zuo_etal_15, Coatman_etal_17, Schulze18, Coffey_etal_19, Marziani_etal_19}.

At redshift larger than 2, both the H$\beta$ line and the MgII line have moved out of the optical observing
window. Such SE estimates have been extrapolated to the CIV $\lambda1549$ emission line in the rest-frame ultraviolet wavelength
\citep{McLure_Jarvis02, Vestergaard02, Vestergaard_Peterson06, Vestergaard_Osmer09, Denney12, Park_etal_13, 
Coatman_etal_16, Coatman_etal_17, Park_etal_17, Sulentic_etal_17, Marziani_etal_19}.

However, given the fact that before the SDSS RM Project \citep[for a technical overview, see][]
{Shen_etal_15}, the RM technique is mainly based on the H$\beta$ emission line for low redshift AGN sample, 
the CIV emission line lacks direct calibrations from large samples. The most recently obtained CIV 
Radius-Luminosity relation based on the SDSS RM Project has raised the number of sources from $\sim$15
to $\sim$67 \citep{Grier_etal_19}. However, it is still controversial whether calibrations 
based on the overlap of the RM and the SE methods can reliably estimate the CIV-based BH masses for high redshift luminous 
quasars \citep{Shen08, Shen_Liu12, Park_etal_13}. 

The CIV emission line is commonly known to show asymmetry and blueshift with respect to the low ionization lines 
\citep{Gaskell_82, Marziani_etal_96, Sulentic_etal_00, Shen08, Richards_etal_11, Denney12, Coatman_etal_16, 
Coatman_etal_17, Sulentic_etal_17, Mejia_etal_18, Vietri_etal_18, Schulze18, GeXue_etal_19, Marziani_etal_19}.
These features suggest that compared with the MgII and H$\beta$ lines, the CIV line width is probably more affected by a 
non-virial velocity component due to disk winds of ejected materials 
\citep{Konigl_Kartje_94, Murray_etal_95, Proga_etal_00, Marziani_etal_10, Richards_etal_11}.

It is therefore essential to consider the effects of these features on the $M_{\rm BH}$ estimates.
One straightforward approach to test the reliability of the CIV-based BH mass estimates is the systematic 
comparison with the Balmer line for the same objects. 

Using a sample of 16 lensed quasars, \cite{Greene10} found no systematic biases in the BH mass estimates between the 
Balmer lines and the CIV line, although the scatter is large. Based on a sample of 12 quasars, \cite{Assef_etal_11} found 
no systematic offsets between the CIV and Balmer line mass estimates, but they did see that the differences between BH 
mass estimates strongly correlate with the logarithm of the ratios of the UV and optical continuum luminosities.

Based on 60 luminous quasars, \cite{Shen_Liu12} found that the CIV line can be calibrated to yield consistent BH
mass estimates with those based on the H$\beta$ line, but the scatter is substantially larger than MgII. They concluded 
that the line width of MgII correlates well with that of H$\beta$ from the SE spectroscopy, while 
the CIV line width is poorly correlated with the MgII or H$\beta$ line widths. Some other studies 
suggested that poor correlations between different line widths play more important roles than the continuum
luminosities in the differences of virial BH mass estimates \citep{Shen08, Denney12, Trakhtenbrot12, Park_etal_13, Runnoe13, Sulentic_etal_17, Coatman_etal_17, Park_etal_17, Mejia_etal_18, Marziani_etal_19}. 

Based on a sample of high-$z$ luminous quasars with $0.9<z<3.1$ and 
$10^{47.4}<L_{\rm bol}<10^{48.4}$ erg s$^{-1}$, \cite{Sulentic_etal_17} confirmed that for high luminosity quasars with strong CIV outflows, the full CIV profile can not perform as a useful virial BH mass estimator
for most quasars. By studying the CIV and Balmer lines of 230 luminous quasars with 
$1.5<z<4.0$ and $10^{45.5}<L_{\rm bol}<10^{48} $ erg s$^{-1}$, 
\cite{Coatman_etal_17} found that with the increase of the CIV line blueshifts, the scatter of the CIV-based BH mass estimates increases dramatically 
compared to the Balmer line-based BH masses, with $\sim$1 dex at the blueshift larger than 5000 km s$^{-1}$ and 
$\sim0.6$ dex at the blueshift around 3000 km s$^{-1}$. With a sample of quasars with $10^{44}<L_{\rm bol}<10^{48.5}$ 
erg s$^{-1}$ and $0<z<3$, \cite{Marziani_etal_19} proposed a scaling law for obtaining the CIV-based BH masses with the 
corrected FWHM of the CIV line. The correction to the CIV FWHM depends on the CIV blueshift and the UV luminosity, and is related to the quasar main sequence (MS).

\cite{Marziani_etal_19} proposed to compare the CIV and H$\beta$ profiles along the quasar MS. In the `Eigenvector 1' 
(EV1) parameter space, the FWHMs of the H$\beta$ broad component (BC) and the REW ratios of the FeII $\lambda4570$ blend to the 
H$\beta$ BC ($R_{\rm FeII}$) are not randomly distributed but instead define a quasar MS 
\citep{Boroson_Green92, Sulentic_etal_00, Marziani_etal_01, Marziani_etal_03, Sulentic_etal_07, Shen_Ho14, Sulentic_etal_17}.
Along with the FWHM of the H$\beta$ BC and $R_{\rm FeII}$, the CIV blueshift as 
one of the other EV1 parameter can be obtained from the rest-frame optical and UV 
spectra, allowing us to understand the relations with CIV blueshifts in the 
context of the EV1 plane. 

In our previous work, we have presented near-infrared (NIR) 
observations of the H$\beta$ and MgII lines for 32 luminous $z\sim 3.5$ quasars with 
$10^{47.5} < L_{\rm bol} < 10^{48.3}$ erg s$^{-1}$ \citep{Zuo_etal_15}. Based on that sample, here we investigate the 
reliability of the CIV-based BH mass for high redshift luminous AGNs. Comparison work based on the sample would complement 
other studies which have proposed empirical corrections to the CIV-based BH masses 
\citep{Assef_etal_11, Denney12, Shen_Liu12, Park_etal_13, Runnoe13, Coatman_etal_17, Sulentic_etal_17, Schulze18, Marziani_etal_19}. 

The dependence of the CIV-based BH mass estimates on the CIV line blueshift and other physical properties will also be 
investigated here. Previous works reported the presence of large CIV blueshift in quasars with the H$\beta$ 
FWHM less than 4000 km/s \citep{Bachev_etal_04, Sulentic_etal_07, Marziani_etal_10}, 
while our work would allow us to extend the detection of large CIV blueshift in quasars with large H$\beta$ FWHM values \citep{Sulentic_etal_17, Vietri_etal_18}.
In addition, the origin of CIV blueshift is assessed using the correlations of blueshift with other quasar properties, 
such as the line width, REW, radio loudness and Eddington ratio ($R_{\rm EDD}$) 
\citep{Richards02, Richards_etal_11, Vietri_etal_18, Sun_etal_18, Marziani_etal_19}. 

We describe our sample and spectral measurements in Section 2. The results are delivered in Section 3 and 
discussed in Section 4. We summarize our main results in Section 5. Throughout the paper, a flat $
\Lambda$CDM cosmology with $\Omega_{\Lambda} = 0.7$, $\Omega_{0} = 0.3$ and $H_{0} = \rm 70\ km\ s^{-1}\ Mpc^{-1}$ is adopted.

\section{Data\label{sec:data}}

\subsection{Quasar Sample\label{subsec:sample}}

We selected our targets from the SDSS DR7 quasar catalog \citep{Schneider10, Shen11} mainly by constraining the redshift and magnitude ranges. To ensure the MgII and H$\beta$ lines residing in NIR spectroscopy, redshifts were restricted between 3.2 and 3.8. Certain redshift ranges were also excluded to avoid the H$\beta$ or MgII lines accidentally falling in telluric absorption bands. Apparent Vega magnitudes in $J$ and $K$ bands were limited to brighter than 17 and 16 magnitudes, respectively.

With these criteria, we selected 32 targets from the main DR7 quasar catalog. Among the 32 targets, 30 were observed with the TripleSpec instrument mounted on the Hale 200 inch telescope, yielding a continuous spectral coverage of 0.95-2.46 $\mu$m simultaneously at a resolution of 2700 \citep{Terry08}. The remaining two objects were observed with the LUCI 1 NIR instrument \citep{Hill12} mounted on the Large Binocular Telescope (LBT). $\ J$ and $K$ bands spectra were obtained with a resolution of 8460 and 6687, respectively. 

The basic data reduction includes flat field correction, background subtraction, wavelength calibration, one dimensional spectra extraction, telluric correction and absolute flux calibration \citep{Cushing04, Becker09, Bian10, Zuo_etal_15}. These reduced spectra were then de-reddened for Galactic extinction \citep{Cardelli89, Schlegel98} and wavelength corrected to the rest frame. We used the H$\beta$ and [OIII] doublets to determine the systemic redshift for each object. The NIR spectra observations and other related details can be found in \cite{Zuo_etal_15}.

After excluding 6 broad absorption line quasars (BALs), there are 2 quasars with NIR spectra labeled as `poor' and 3 quasars with NIR spectra labeled as `median'. Finally, 21 quasars with NIR spectra labeled as `good' are left \citep{Zuo_etal_15}. Among the 21 quasars, there are 20 quasars with full coverage of the H$\beta$ line, 20 quasars with full coverage of the MgII line and 19 quasars with full coverage of the H$\beta$ line and MgII line. All of them have good SDSS spectra with full coverage of the CIV line  (signal-to-noise ratio per spectral resolution elements larger than 10). Thus, these 19 targets are adopted for the following analysis in this work.

The optical spectra are all collected from the SDSS DR14 database\footnote[9]{https://dr14.sdss.org/optical/spectrum/search}. Table \ref{table:spectra_dr7and14} lists the number of the SDSS spectroscopic observations for each quasar, the related information from \cite{Shen11} (SDSS DR7) and the SDSS spectrum with the highest S/N per pixel obtained after the SDSS DR7 (SDSS DR7$+$). For the 10 quasars with SDSS DR7$+$ spectra, the mean S/N of the SDSS DR7 spectra and that of SDSS DR7$+$ spectra are 25.3 and 32.5, respectively. Our analysis are based on the spectrum with the highest S/N for each quasar, i.e., the SDSS DR7 spectra for 9 quasars and the SDSS DR7$+$ spectra for 10 quasars.
 
\subsection{Spectral Measurements}\label{subsec:spec_measurement}

The procedures employed to derive the properties of the emission lines in the NIR spectra 
(e.g. H$\beta$, [OIII] and MgII) were described in detail in \cite{Zuo_etal_15}. The line properties and 
the continuum luminosities around these emission lines are taken directly from \cite{Zuo_etal_15}. 
Here we briefly review the procedure to measure the NIR spectra and present our new measurements from 
the optical spectra (particularly the CIV line) in detail.

For each emission line, we locally fit a pseudo-continuum to the continuum dominated wavelength range 
around the line. The pseudo-continuum consists of a power-law continuum and Fe II emissions.
As mentioned in earlier studies, the contribution from Fe II around the CIV line 
is quite small \citep{Shen11, Trakhtenbrot12}. Considering the difficulty of constraining 
the Fe II features, including Fe II in the emission line fitting will introduce extra uncertainties. 
Therefore we decide not to include the Fe II features in the CIV line fitting. The 
power-law continuum fitting windows around the CIV line are generally selected to be 
[1445, 1465] \AA\ and [1700, 1705] \AA. To alleviate the effects of narrow absorption lines, 
during the pseudo-continuum fitting and subsequent emission-line fitting, we rejected data points that 
are 5$\sigma$ below the 20-pixel boxcar-smoothed spectrum.

After subtracting the pseudo-continuum, the emission lines were fitted with multiple Gaussians 
\citep[for more details, see][]{Shen11, Zuo_etal_15, Shen_etal_19}. In the wavelength range from 
4700 to 5100 \AA, we fitted the line profiles with 5 Gaussians: 2 for the BC of 
the H$\beta$ line, 1 for the narrow component (NC) of the H$\beta$ line and 2 Gaussians for the [OIII] 
$\lambda\lambda4959$,$5007$ doublets. 
Each Gaussian fitted to the H$\beta$ BC was generally constrained with the line center offset 
($\Delta \log(\lambda_{\rm rf}$ (\AA))) less than 0.015 and the FWHM less than 35250 km s$^{-1}$. 
Minor modifications of the fitting parameters were made if necessary.
The NC of H$\beta$ and the [OIII] doublets were tied together 
with the same line center offsets ($\Delta \log(\lambda_{\rm rf}$ (\AA)) $<$ 0.005) from their 
theoretic values and the same FWHM. The upper FWHM limit of 
the NC was imposed as 1200 km/s \citep{Shen11}. If needed, we introduced two additional Gaussians with the 
same FWHM for the extended wings of the [OIII] doublets, which were not tied to the NC of H$\beta$ 
\citep{Vietri_etal_18}. Among the 19 quasars, the [OIII] doublets of 7 quasars (listed in bold 
in Table \ref{table:civdetails}) were modeled with two pairs of Gaussians \citep{Zuo_etal_15}. 

In the wavelength range from 2700 to 2900 \AA, the BC and NC of the MgII line were modeled with 2 
Gaussians and 1 Gaussian, respectively. Each Gaussian fitted to the MgII BC was generally constrained 
with $\Delta \log(\lambda_{\rm rf}$ (\AA))) less than 0.015 and the FWHM less than 35250 km s$^{-1}$. 
The line center and the FWHM of the MgII NC were tied to that of the H$\beta$ NC.

The pseudo-continuum subtracted spectra at 1500 $< \lambda < 1700$ \AA\ are also modeled with
multiple Gaussians. Since the existence of a strong NC in the CIV line is still 
controversial, which may be difficult to disentangle in the spectra, first we use Model A to fit 
the region around the CIV line: 3 Gaussians for the BC of CIV, 1 for the BC of the HeII $\lambda1640\ $ 
line and 1 for the BC of the OIII] $\lambda1663$ line. Each Gaussian fitted to the CIV/HeII/OIII] 
BC is constrained with $\Delta \log(\lambda_{\rm rf}$ (\AA))) less than 0.015/0.008/0.008 and the FWHM 
less than 35250/14100/14100 km s$^{-1}$.
Any Gaussian component contributing less than 5\% of the total flux is rejected when estimating the FWHMs 
of the BC of the CIV line.

We then use Model B to fit the CIV line complex: 2/1 Gaussians for the BC/NC of CIV, 
1/1 for the BC/NC of the HeII line and 1/1 for the BC/NC of the OIII] line.  
The FWHM and the line center of the NC of these three lines are tied together.
Here, the upper FWHM limit of the CIV NC is relaxed to 1600 km s$^{-1}$.

Based on a joint analysis of the reduced $\chi^2$ of line fitting from the two models and visual inspection,
we identify 7 quasars ($J015741.57-010629.6$, $J025021.76-075749.9$, $J025905.63+001121.9$, 
$J030449.85-000813.4$, $J075303.34+423130.8$, $J080819.69+373047.3$ and $J090033.50+421547.0$) 
with a NC for their CIV lines and adopt the fitting results from Model B. 
For the other targets, we use Model A to fit the CIV line. The comparisons are 
presented in Appendix. 

The spectral fitting results for the wavelength range of 1500-1700 \AA\ are shown in 
Fig. \ref{fig:civspec_all}. 
Table \ref{table:civdetails} lists the line shift ($\Delta V$), REW and 
FWHM of the three Gaussians for the CIV line of all the 19 quasars. Note that each Gaussian fit to the CIV 
line may not have a robust physical interpretation.

In Table \ref{table:civdetails}, $\Delta V_{\rm i}$ is the line shift of the $i$-th Gaussian profile, which is calculated 
by comparing the fitted line center ($\lambda_{\rm i}$) with the expected rest-frame wavelength of CIV $\lambda1549$ according to 
[OIII] $\lambda 5007$ ($\lambda_{\rm lab}$), 
\begin{equation}
\Delta V_{\rm} = \frac{\lambda_{\rm lab} - \lambda_{\rm i}}{\lambda_{\rm lab}} \times c
\end{equation}
\begin{equation}
\lambda_{\rm lab} = 1549.06 \times \frac{\lambda_{0} (\rm [OIII] \lambda5007)}{5008.24} 
\end{equation}  
where $\lambda_{0}$([OIII] $\lambda5007)$ is the peak wavelength of the first Gaussian fitting to the 
[OIII] $\lambda5007$ and $c$ is the speed of light in a vacuum. 1549.06 \AA\ and 5008.24 \AA\ are the average rest-frame wavelength of the 
unsaturated CIV $\lambda\lambda1548.2,1550.8$ doublets and the rest-frame wavelength 
of the $\rm [OIII] \lambda5007$ line, respectively \citep{Berk_etal_01}. In each panel of Fig. \ref{fig:civspec_all}, 
the vertical red dashed line refers to $\lambda_{\rm lab}$.

Observations show that even under the simple emission situation in a planetary 
nebulae, the CIV doublet are close to being saturated with the intensity ratio $\sim$0.8-2.0 \citep[e.g.][]{Feibelman_83}. 
In that case, assuming equal contribution from both components of 
the CIV doublet, the average rest-frame wavelength of CIV is 1549.48 \AA. However, to maintain the consistency with a 
number of previous works \citep[e.g.][]{Berk_etal_01, Shen11, Shen_Liu12}, the average rest-frame wavelength of 
the CIV doublet is adopted as 1549.06 \AA\ under the assumption of the unsaturated CIV doublets with the intensity ratio 
as 2, though it would generally over-estimate the CIV blueshift by $\sim$80 km s$^{-1}$. 

The line shift $\Delta V_{\rm CIV}$ of the CIV BC is calculated with two methods. In the first method, 
$\Delta V^{\rm peak}_{\rm CIV}$ is calculated with the peak wavelength of the best-fitting BC ($\lambda_{0}$), as
shown with the vertical black dashed line in each panel of Fig. \ref{fig:civspec_all}.
While in the second method, $\Delta V^{\rm half}_{\rm CIV}$ is calculated with the wavelength that bisects the 
commulative total flux of the best-fitting BC ($\lambda_{\rm half}$) as
\begin{equation}  
\Delta V^{\rm half}_{\rm CIV} = \frac{\lambda_{\rm lab} - \lambda_{\rm half}}{\lambda_{\rm lab}} \times c
\end{equation}
The vertical black long dashed line in each panel of Fig. \ref{fig:civspec_all} refers to $\lambda_{\rm half}$.

The line shifts calculated with both methods are listed in Table~\ref{table:civdetails}. 
Compared with $\Delta V^{\rm peak}_{\rm CIV}$, $\Delta V^{\rm half}_{\rm CIV}$ yields a more straightforward way to measure the 
CIV line shift and is generally subjected to a smaller uncertainty.  
In our following analysis, the calculated $\Delta V^{\rm half}_{\rm CIV}$ values are adopted as the CIV 
line shifts that are abbreviated as $\Delta V_{\rm CIV}$ for simplicity.

We also measure the CIV emission line asymmetry ($\rm AS_{\rm CIV}$) as the ratio of the widths red and blue of the line 
centroid from the model fitting of the CIV BC:
\begin{equation}
\rm AS = \ln \frac{\lambda_{red}}{\lambda_{0}} / \ln \frac{\lambda_{blue}}{\lambda_{0}}
\end{equation}
where $\lambda_{\rm red}$ and $\lambda_{\rm blue}$ are the wavelength at half peak flux red and blue of the line centroid  
\citep{Shen_Liu12}, as shown with the vertical blue dashed lines from right to left in each panel of 
Fig. \ref{fig:civspec_all}.

\begin{figure*}
\plotone{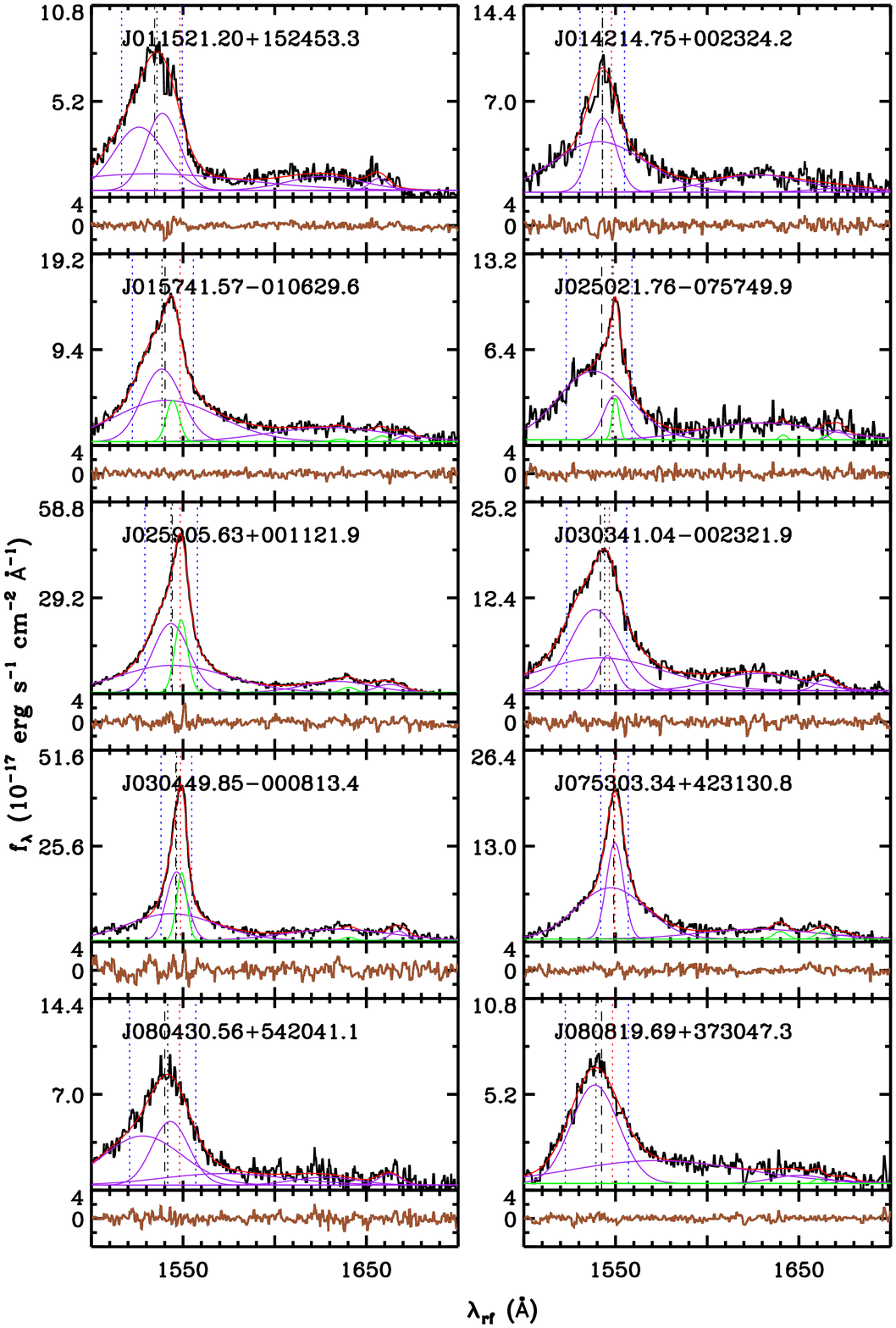}
\caption{Fitting results of the CIV line complex in the wavelength range of 1500-1700 \AA\  for the 19 quasars, 
where the spectrum in each panel is shown in black, the combined model fitting is shown in red, the individual 
Gaussian for the BC is shown in purple, the individual Gaussian for the NC is shown in green and the fitting
residuals are shown in brown. The vertical blue dashed lines from left to right refer to the wavelengths of 
$\lambda_{\rm blue}$ and $\lambda_{\rm red}$. The vertical red dashed line refers to the wavelength of $\lambda_{\rm lab}$. 
The vertical black dashed and long dashed lines refer to the wavelengths of $\lambda_{0}$ and $\lambda_{\rm half}$, respectively.
Except 7 quasars ($J015741.57-010629.6$, $J025021.76-075749.9$, $J025905.63+001121.9$, 
$J030449.85-000813.4$, $J075303.34+423130.8$, $J080819.69+373047.3$ and $J090033.50+421547.0$), the CIV line of the other quasars is not fit with a NC.}
\label{fig:civspec_all}
\end{figure*}
\renewcommand{\thefigure}{\arabic{figure}(Cont.)}
\addtocounter{figure}{-1}
\begin{figure*}
\plotone{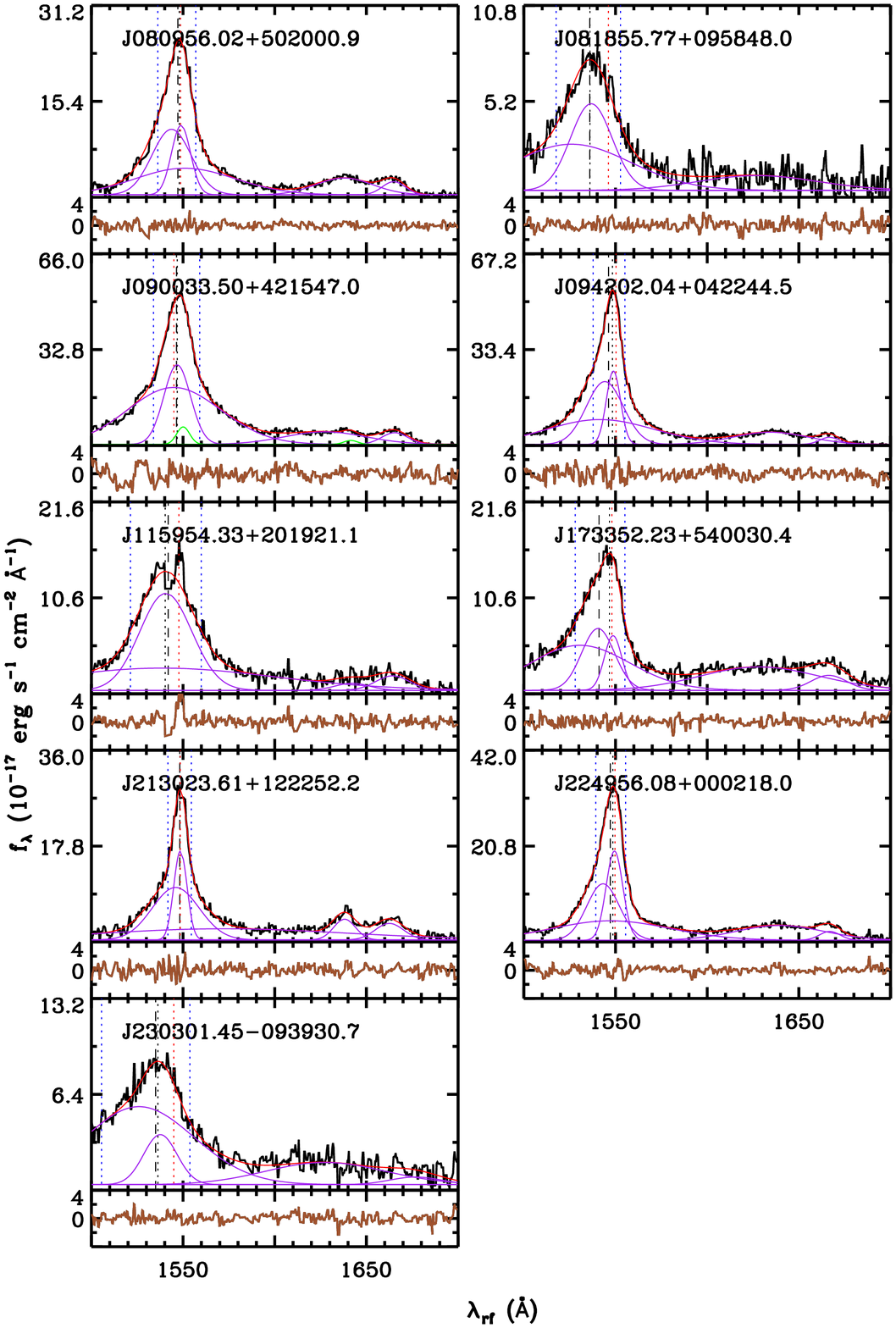}
\caption{(Continued)}
\end{figure*}
\renewcommand{\thefigure}{\arabic{figure}}

\subsection{Uncertainty Estimation}
We apply the Monte Carlo approach to estimate the uncertainties of the fitting parameters 
\citep{Shen11, Shen_Liu12}. For each object, 
100 random mock spectra are created by introducing random Gaussian noises to the original spectrum
using the flux density errors of the original spectrum. 
We then fit the mock spectra with the same fitting strategy. The 1 $\sigma$ dispersion centered on the 
median of these measurements is taken as the uncertainty, which accounts for the statistical 
uncertainties and the systematic uncertainties due to the flux errors and the ambiguities in multi-component 
spectral fitting, respectively.

Based on the quasars with multiple-epoch observations obtained from the SDSS-RM project, \cite{Sun_etal_18} 
justified the uncertainty estimates by exploring the distributions of quasar properties between close-epoch
(i.e., rest-frame time interval < 2 days) pairs. Taking these as the true uncertainties, the uncertainties 
estimated using the same Monte Carlo approach are smaller by a factor 
of $\sim$ 1.2 to 1.7. Therefore, the uncertainties obtained from the Monto Carlo approach should 
be scaled up by a factor of $\sim$ 1.2 to 1.7.

To estimate the errors caused by the positioning of the continuum, after globally fitting the 
pseudo-continuum underlying H$\beta$ and MgII of the 19 objects, we calculate the difference between the 
obtained parameters and the corresponding parameters obtained with the method in 
Section~\ref{subsec:spec_measurement}.
Given the median absolute difference of $L_{3000}$ and $L_{5100}$ as $\sim$0.012 and 0.043 dex, respectively,  
the larger one (0.043 dex) can be taken as a rough estimate of the error of the continuum luminosity caused by the 
positioning of the continuum. In the same way, the error of the line FWHM caused by the positioning of the 
continuum is $\sim$ 410 km s$^{-1}$. 
These estimated errors can be propagated into the statistical uncertainties
to account for the true errors of the continuum luminosities and the line FWHMs.  

\section{Results\label{sec:results}}
\subsection{Comparison with the H$\beta$-based BH Mass Estimates\label{subsec:results-comparison}}
\begin{figure*}
\plotone{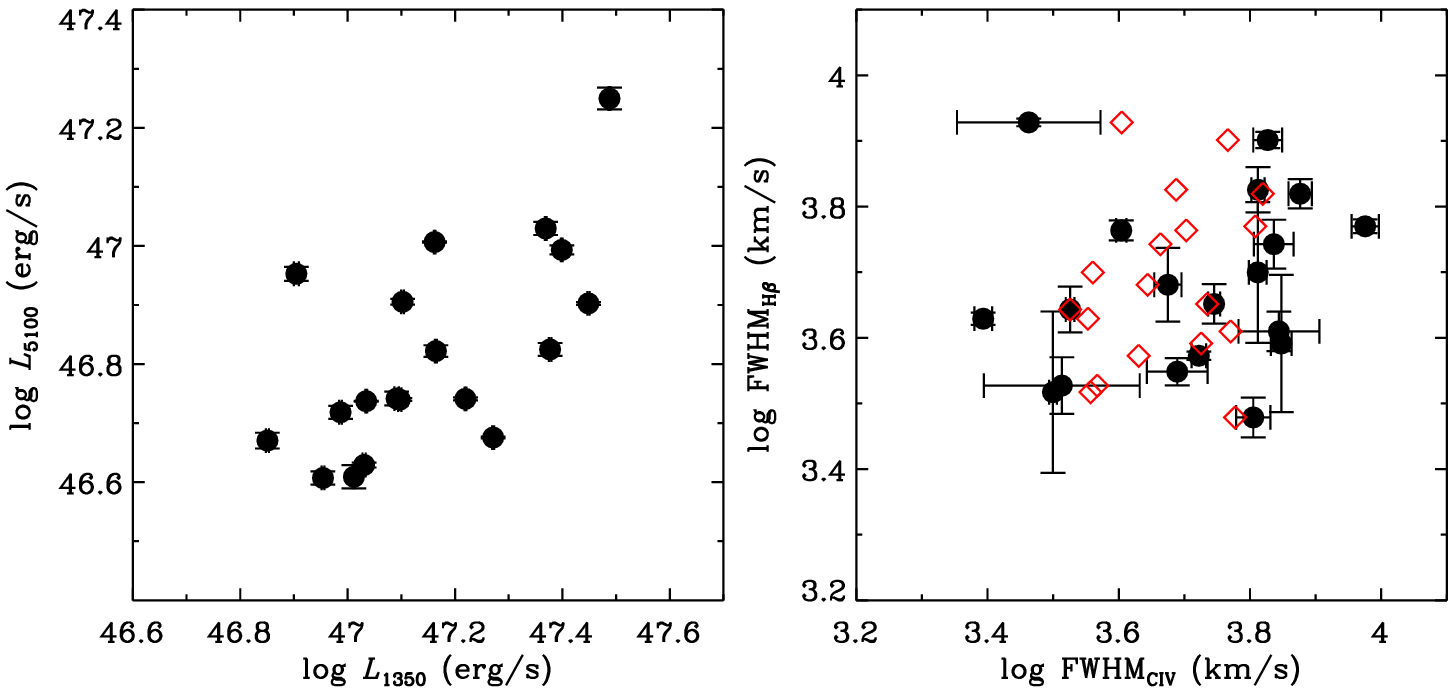}
\caption{Left panel: Correlation between $\log \ L_{1350}$ and $\log \ L_{5100}$. Right panel: Correlation between 
$\log\ \rm FWHM_{H\beta}$ and $\log\ \rm FWHM_{CIV}$, where the red diamonds refer to the corrected 
$\rm log\ FWHM_{CIV}$ using Eq.~\ref{eq:fwhmcor}. More details can be found in Section\ 
\ref{subsec:results-corrected_fwhm}.}
\label{fig:l_fwhm_hbciv}
\end{figure*}

The continuum luminosities and the CIV line properties of the 19 targets are tabulated in 
Table \ref{table:continuum-line-params}. We use the Spearman's rank correlation coefficient ($r$) to 
describe the monotonic correlation between the continuum luminosity at 1350 \AA\ ($L_{1350}$) and 
the continuum luminosity at 5100 \AA\ ($L_{5100}$). As shown in the left panel of Fig. \ref{fig:l_fwhm_hbciv}, 
we find a strong correlation between $\ L_{1350}$ and $\ L_{5100}$ with $r\sim0.64$ at a confidence level 
over 99\%. The slope from the bisector linear regression fitting using the BCES estimator 
\citep{Akritas_Bershady96} is $1.1\pm0.4$, consistent with that in \cite{Shen_Liu12}.
As shown in the right panel of Fig. \ref{fig:l_fwhm_hbciv}, there is a poor
correlation between the CIV FWHM and the H$\beta$ FWHM with $r \sim 0.28$ ($p\sim0.25$), 
suggesting that the clouds emitting the two features
do not fully share the same velocity distribution.

The virial BH mass estimates are expressed as follows:
\begin{equation} \label{eq:mbh}
\log(\frac{M_{\rm BH, vir}}{M_{\odot}}) = a + b\ \log (\frac{L}{10^{44} {\rm erg\ s^{-1}}}) + c\ \log {\rm (\frac{FWHM}{km\ s^{-1}})},
\end{equation}
where $L$ and FWHM are the continuum (line) luminosity and the width of one emission line, respectively. 

As mentioned in \cite{Zuo_etal_15}, the H$\beta$-based BH masses are calculated using the calibrations from \cite{Vestergaard_Peterson06}, 
with $(a$, $b$, $c) = ($0.91, 0.50, 2.00$)$. The MgII-based BH masses are also considered here for comparison, 
with $(a$, $b$, $c) = ($1.07, 0.48, 2.00$)\ $\citep{Zuo_etal_15}. 

The CIV-based BH masses are estimated using the calibration from \cite{Vestergaard_Peterson06}, with $(a$, $b$, $c)= 
($0.66, 0.53, 2.00$)$. 
The errors of the CIV-based BH masses listed in Table~\ref{table:continuum-line-params} are estimated as the 1 $\sigma$ dispersion 
centered on the median of the measurements from the Monte Carlo approach, not including the intrinsic error of the BH mass 
single-epoch virial relation ($\sim0.4$ dex) \citep{Vestergaard02, Vestergaard_Peterson06, Onken04}.

\begin{figure}
\plotone{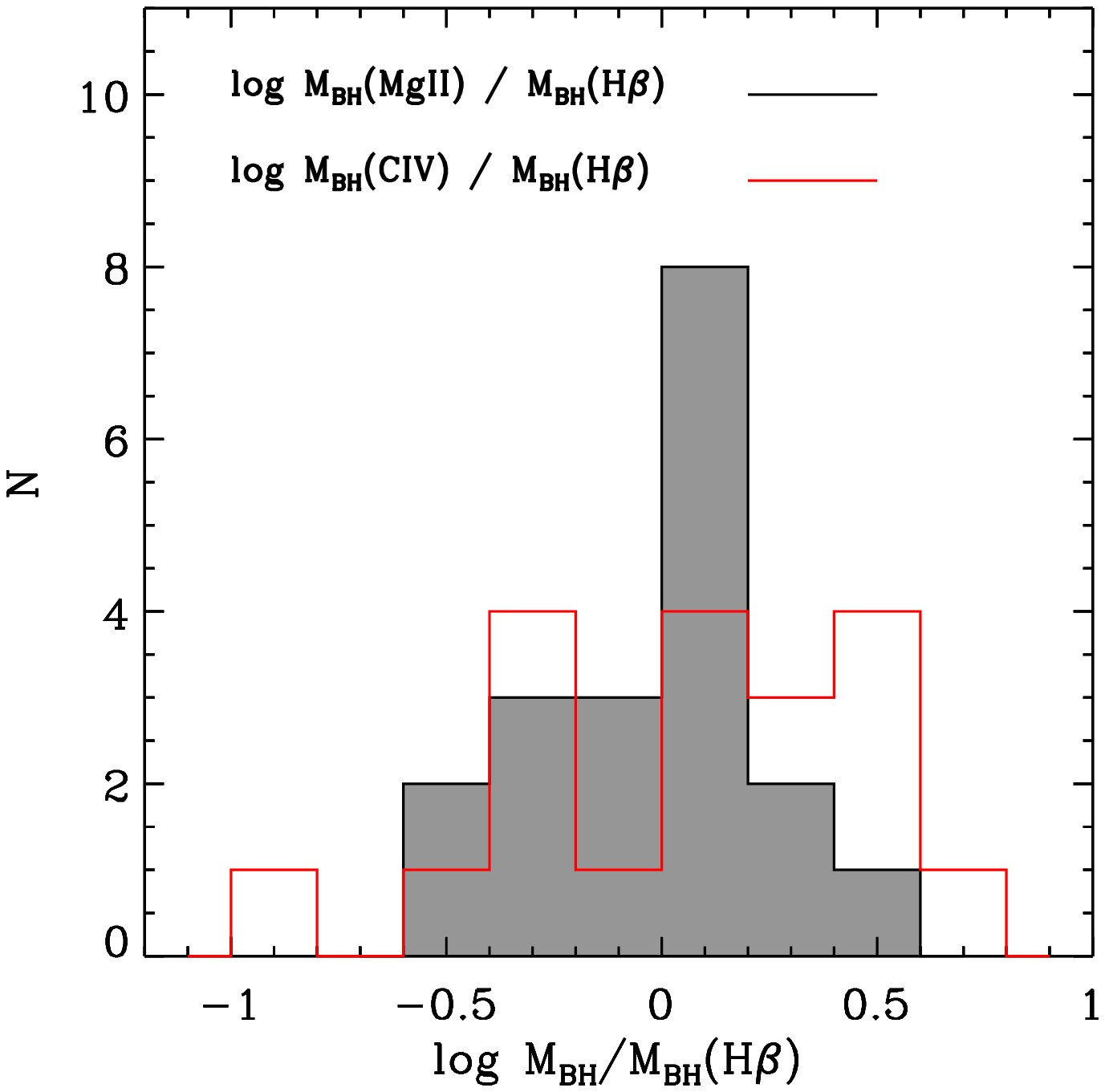}
\caption{Distributions of $\log\ \rm M_{BH}(CIV) / M_{BH} (H\beta)$ (red histogram) and $\log\ \rm M_{BH}(MgII) / M_{BH}(H\beta)$ 
(grey shaded histogram).
\label{fig:deltambh}}
\end{figure}

Taking the H$\beta$-based BH mass estimates as the reference values, we compare the CIV-based 
and the MgII-based BH masses with the reference values in \cite{Zuo_etal_15}. The histograms of the logarithm of the CIV- 
and MgII-based BH mass ratios, i.e., 
$\log M_{\rm BH}({\rm CIV}) / M_{\rm BH}({\rm H}\beta)$ and $\log M_{\rm BH}({\rm MgII}) / M_{\rm BH}({\rm H}\beta)$, 
are shown in Fig. \ref{fig:deltambh}. The values of $\log M_{\rm BH}({\rm CIV}) / M_{\rm BH}({\rm H}\beta)$ are between 
$-0.85$ dex and $0.67$ dex, with a median value of $0.110\pm0.647$ dex. 
The median value of $\log M_{\rm BH}({\rm MgII}) / M_{\rm BH}({\rm H}\beta)$ is $0.041\pm0.394$ dex. 
Here the scatters are calculated as the inner 50 percentile of the distributions of the logarithm of the mass ratios. 

Considering the intrinsic error of the SE virial BH mass estimates as $\sim0.4$ dex,
all the quasars show consistent MgII-based BH mass estimates with the corresponding H$\beta$-based 
BH masses, while 63\% quasars show consistent CIV-based BH mass estimates with the H$\beta$-based $M_{\rm BH}$ values. The BH mass differences of the 
other 37\% quasars suggest that the CIV-based BH mass estimates still need to be corrected to better match the H$\beta$-based 
BH masses.

As shown in Table \ref{table:spectra_dr7and14}, for each quasar in our sample the median S/N per pixel of the 
SDSS spectra is larger than 15. For the NIR spectrum of each quasar, the median S/N per spectral resolution 
element of 3 pixels is no less than 10. 
Using the SDSS DR7 sample, \cite{Shen11} found that the bias affecting the continuum luminosities and FWHMs 
during the spectral measurements are negligible even if S/N is reduced to as low as $\sim$ 5. This is further confirmed 
in \cite{Runnoe13}. Given the relatively high S/N of the NIR and optical spectra in our sample, we argue that the difference of 
the S/N of SDSS optical spectra and NIR spectra of our quasars makes small contribution to the differences between the CIV- 
and H$\beta$-based BH masses.

The continuum variation is typically at the level of $\sim$ 0.1 mag for average SDSS quasars 
\citep{MacLeod_etal_2012, Zuo_etal_2012}. Since the luminosity enters into the BH mass estimates as 
the square-root, the luminosity uncertainty does not make a large contribution to the BH mass estimates 
\citep{Coatman_etal_17}. For the line shape variability, \cite{Wilhite_etal_2007} found that the variation 
of the CIV line FWHM is less than 0.05 dex, using 615 high z quasars with spectra observed at two epochs. 
They concluded that the inherent continuum and line shape variability contributes $\sim 20\%$ to the 
BH mass variations ($\sim$ 0.08 dex) between different epochs. 

As listed in Table 1, the time differences between the Near-IR spectroscopy and the SDSS spectroscopy 
range from 49 days to 12 years, corresponding to 11 days-2.6 years in the rest frame of quasars
($\Delta T_{\rm rf}$), where 8 out of the 19 quasars were observed with time differences larger than 2 years 
in the rest frame of quasars. The absolute logarithm of the median BH mass ratio 
($\log M_{\rm BH}({\rm CIV}) / M_{\rm BH}({\rm H}\beta)$) for quasars with $\Delta T_{\rm rf} < 2$ 
years is measured to be 0.11$\pm$0.46 dex, while for quasars with $\Delta T_{\rm rf} > 2$ years, the 
absolute logarithm of the median BH mass ratio is 0.43$\pm$0.53 dex. 

However, we find that the median $\log M_{\rm BH}({\rm H\beta})$ value for quasars with 
$\Delta T_{\rm rf}<2$ years is larger by 0.23 dex than that for quasars with $\Delta T_{\rm rf} > 2$ years, 
and the median $\log M_{\rm BH}({\rm CIV})$ value for quasars with $\Delta T_{\rm rf}<2$ years is smaller by 
0.08 dex than that for quasars with $\Delta T_{\rm rf} > 2$ years. 
That means the relative larger 
$\lvert \log M_{\rm BH}({\rm CIV}) / M_{\rm BH}({\rm H}\beta) \rvert$ with $\Delta T_{\rm rf} > 2$ years
is mainly due to the distributions of the CIV- and H$\beta$-based BH masses with the observing time 
difference, which are not related to the quasar intrinsic properties.

\subsection{Dependences of the Mass Differences on Different Parameters\label{subsec:results-mass_ratio}}
To investigate whether the logarithm  of the mass ratios depend on the CIV emission line properties, 
in Eq.~\ref{eq:mbh} we assume that there is no dependence on the FWHM by fixing $c$ as 0 and then calibrate 
the CIV-based BH masses to the reference H$\beta$-based BH masses using the LINMIX$\_$ERR approach \citep{Kelly07}. If $b$ is 
adopted as 0.530 \citep{Vestergaard_Peterson06}, the best-fitting result yields the $a$ value as 8.01$\pm$0.26. 
The median logarithm of the mass ratio is 0.025$\pm$0.360.

To investigate the dependence of the mass ratios on the continuum luminosities, in Eq.~\ref{eq:mbh} we 
assume that there is no dependence on $L$ by fixing $b$ as 0 and then calibrate the CIV-based BH masses 
to the reference BH masses using the LINMIX$\_$ERR approach \citep{Kelly07}. If $c$ is adopted as 2.00
 \citep{Vestergaard_Peterson06}, the best-fitting $a$ value is 2.26$\pm$0.38. The median logarithm 
of the mass ratio is 0.006$\pm$0.432. 

The median logarithm of the mass ratios and fitting results are tabulated in Table \ref{deltambh}. 
We find that under the aforementioned assumptions, the scatters are large. Moreover, the scatter is smaller when the dependence of the BH mass estimates on the FWHM is assumed to be zero. It indicates that the scatter of 
the mass differences is more related to the line properties than the continuum luminosities.

For the 19 targets, we further investigate the Spearman rank correlations between the logarithm of 
the mass ratios $\log M_{\rm BH}({\rm CIV}) / M_{\rm BH} ({\rm H}\beta)$ and detailed spectral properties, 
as tabulated in Table  \ref{residual}. There is a strong correlation with 
$\rm FWHM_{CIV} / FWHM_{H\beta}$ ($r\sim$0.98, $p<0.01$),  $\rm FWHM_{CIV}$ ($r\sim$0.67, $p < 0.01$) 
, a moderate correlation with the CIV blueshift $\Delta V_{\rm CIV}$ ($r\sim$0.54, $p \sim 0.02$) 
and a moderate anti-correlation with the CIV asymmetry AS$_{\rm CIV}$ ($r\sim-0.62$, $p<0.01$).
However, we note that the strong relation with $\rm FWHM_{CIV} / FWHM_{H\beta}$ is simply due to the 
fact that the calculated $M_{\rm BH} ({\rm CIV}) / M_{\rm BH} ({\rm H\beta})$ values are proportional 
to $(\rm FWHM_{CIV} / FWHM_{H\beta})^2$. No significant correlations with other parameters is found, 
such as the luminosity or the logarithm of the luminosity ratio.

\begin{figure*}
\plotone{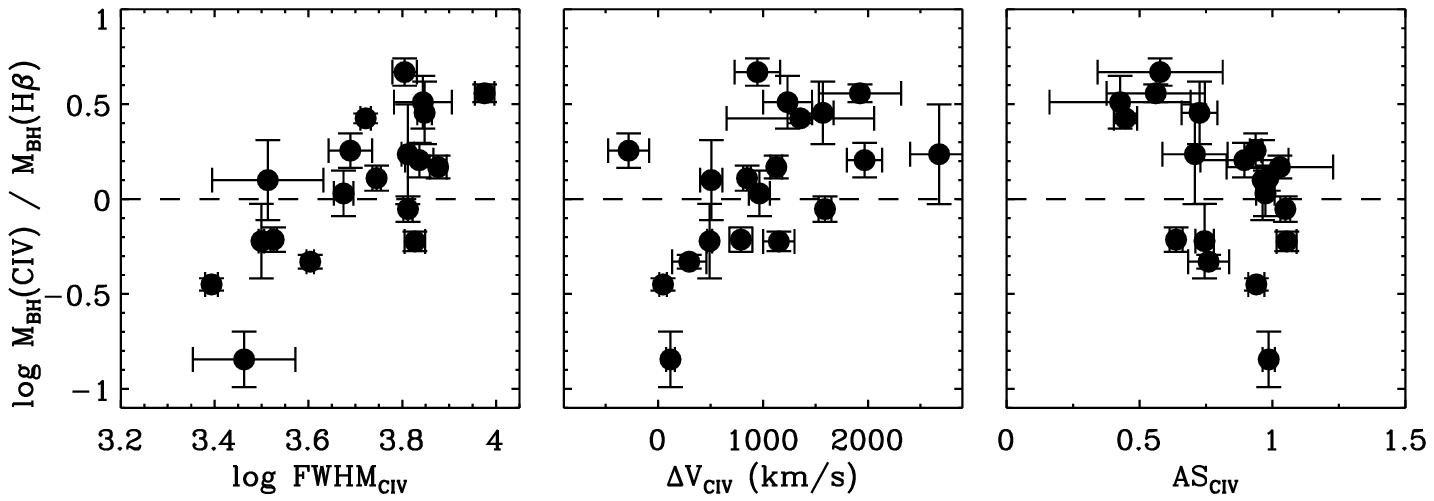}
\caption{Dependence of the logarithm of the CIV-to-H$\beta$ BH mass ratios on $\rm FWHM_{CIV}$ (left), the  CIV 
  blueshifts (middle) and the CIV asymmetries (right). The dashed line in each panel refers to the situation where the CIV- 
  and H$\beta$-based BH masses are equal. 
\label{fig:mbhratio_all}}
\end{figure*}

It suggests that the logarithm of the mass ratio is mainly affected by the line properties, such as 
FWHM$_{\rm CIV}$, $\Delta V_{\rm CIV}$ and AS$_{\rm CIV}$. The dependence of $\log M_{\rm BH}({\rm CIV})/M_{\rm BH}({\rm H}\beta)$ on 
the observed parameters is shown in Fig. \ref{fig:mbhratio_all}, including the dependences on 
$\rm FWHM_{CIV}$, $\Delta V_{\rm CIV}$ and AS$_{\rm CIV}$. Influences caused by line properties, such as the 
blueshift and asymmetry could be taken into account to further reduce the difference between the CIV- and H$\beta$-
based BH mass estimates.

\subsection{Correlation between the CIV FWHM and the CIV Blueshift \label{subsec:results-correlation_fwhm_shift}}
\begin{figure}
\plotone{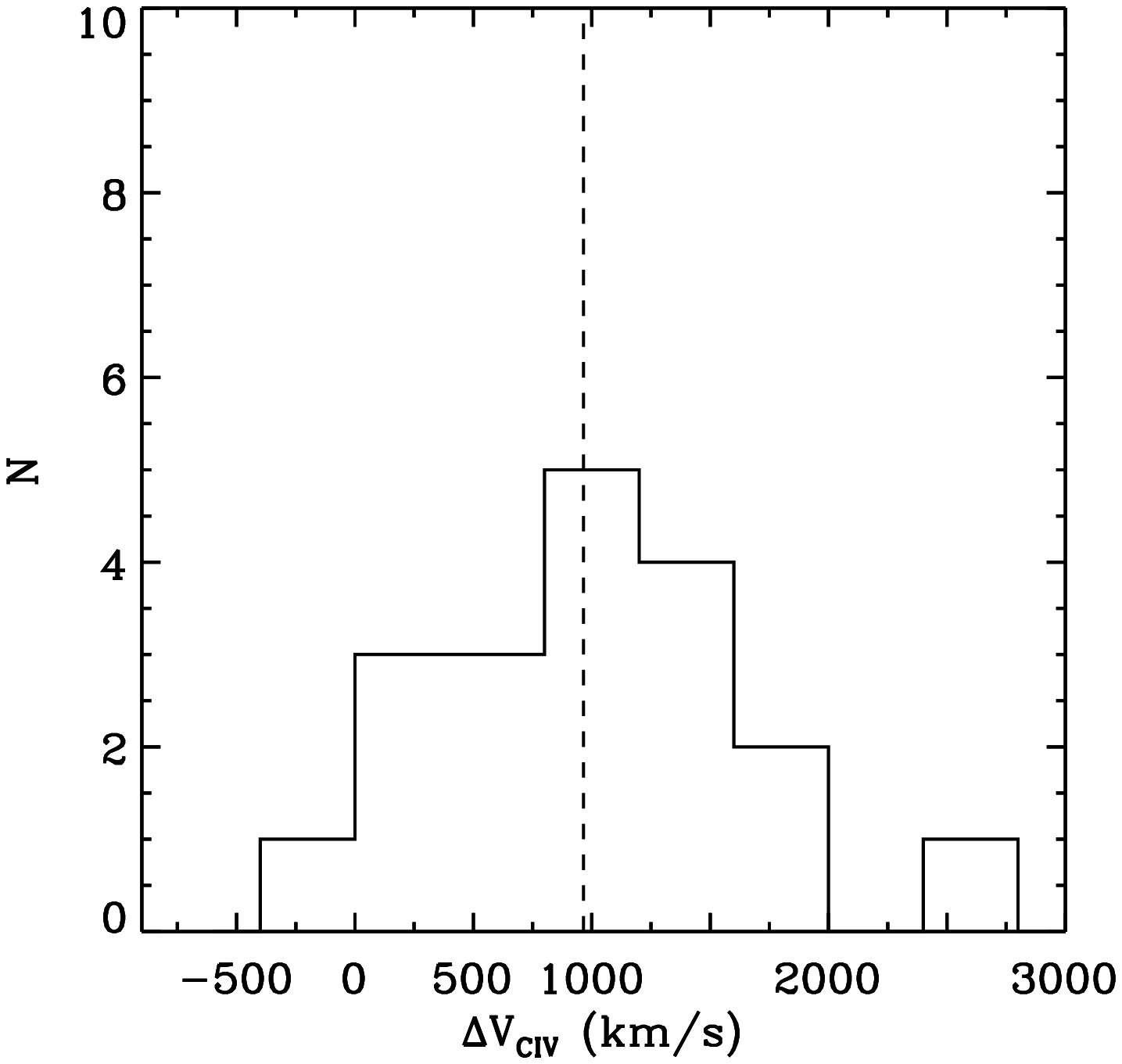}
\caption{Distribution of the CIV emission line blueshifts with respect to [OIII] $\lambda5007$.\label{fig:civshift_distr}}
\end{figure}

The distribution of the derived CIV blueshifts is shown in Fig. \ref{fig:civshift_distr}. Among the 19 quasars, 18 quasars exhibit positive CIV blueshifts, with the median value as 1126 km s$^{-1}$ and the inner 50 percentile of the distribution as 1064 km s$^{-1}$. 
14/19 of the quasars show CIV blueshifts larger than 500 km s$^{-1}$.
Uncertainties of the CIV blueshifts range from 25 to 703 km s$^{-1}$, with the median value at 110 km s$^{-1}$. 
The uncertainties are generally larger for larger blueshifts.

\begin{figure}
\plotone{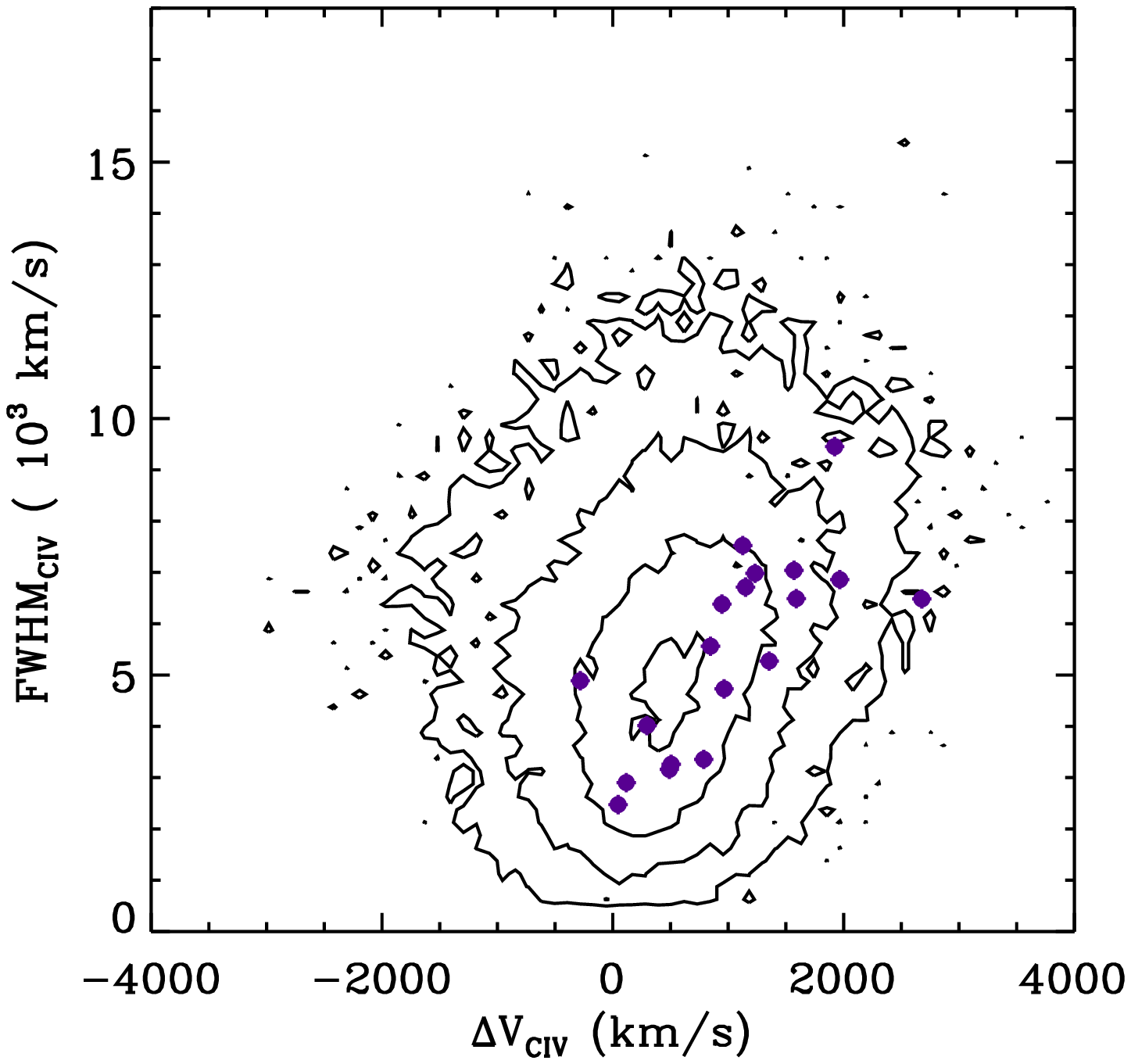}
\caption{Correlation between $\rm FWHM_{CIV}$ and the CIV blueshifts for the 18 quasars
with positive $\Delta V_{\rm CIV}$ values. The contours are the analogue data from the 
large DR7 quasar sample of 44426 quasars with the CIV FWHW values and the CIV blueshifts relative to 
the broad MgII line. The lines refer to the 25, 50, 75, 95 percent contours centered at the 
maximum probability point.
\label{fig:civshift_fwhmciv}}
\end{figure}

As shown in Fig. \ref{fig:civshift_fwhmciv}, for our sample with $L_{\rm bol} \sim 10^{47.5} - 10^{48.3}$ erg s$^{-1}$, the CIV 
FWHM strongly correlates with the CIV blueshift, with the Spearman correlation coefficient as 0.78 ($p < 0.01$). 
Consistent results were found in many previous studies (e.g., Richards et al. 2002; Sulentic et al. 2007; 
Shen\&Liu 2012; Coatman et al. 2016; Sun et al. 2018; Vietri et al. 2018; Marziani et al. 2019), suggesting
that the CIV FWHM is likely a combination of a virialized component and an outflow component.     

For the large DR7 quasar sample, the CIV blueshifts relative to the MgII line can be estimated from the catalogued 
velocity shifts of CIV and the velocity shifts of broad MgII relative to the systematic redshifts \citep{Schneider10, Shen11}.
For a better comparison, we superimpose analogue data in contours, i.e., the CIV blueshifts relative to the broad MgII 
line and the CIV line FWHM, from the large DR7 quasar sample of 44426 quasars with $1.5<z<5.0$ located in the FIRST footprint with 
radio loudness less than 10, non-zero REW$_{\rm CIV}$, FWHM$_{\rm CIV}$, $\log L_{1350}$ and $\log R_{\rm EDD}$
values \citep{Shen11}. The radio loudness ($R$) is defined as the ratio of the flux density at rest-frame 6 cm to the 
flux density at 2500 \AA\ \citep{Jiang2007, Shen11}. The border lines represent the 25, 50, 75, 95 percent contours 
centered at the maximum probability point. Similar to \cite{Coatman_etal_16}, we find that the FWHMs of quasars with large CIV blueshifts ($\approx$ 1500 km/s) are 
about 2 times higher than those with moderate blueshifts ($\approx$ 300 km/s). 

\subsection{Corrections of the CIV FWHM \label{subsec:results-corrected_fwhm}}
Given the correlation between FWHM$_{\rm CIV} /$ FWHM$_{\rm H\beta}$ and the CIV blueshift ($r\sim056$, $p\sim0.01$), 
${\rm FWHM_{CIV}}$ can be calibrated with the CIV blueshift to get better agreement with ${\rm FWHM_{H\beta}}$, which in turn 
results in a slightly more accurate BH mass estimate.

\begin{figure*}
\plotone{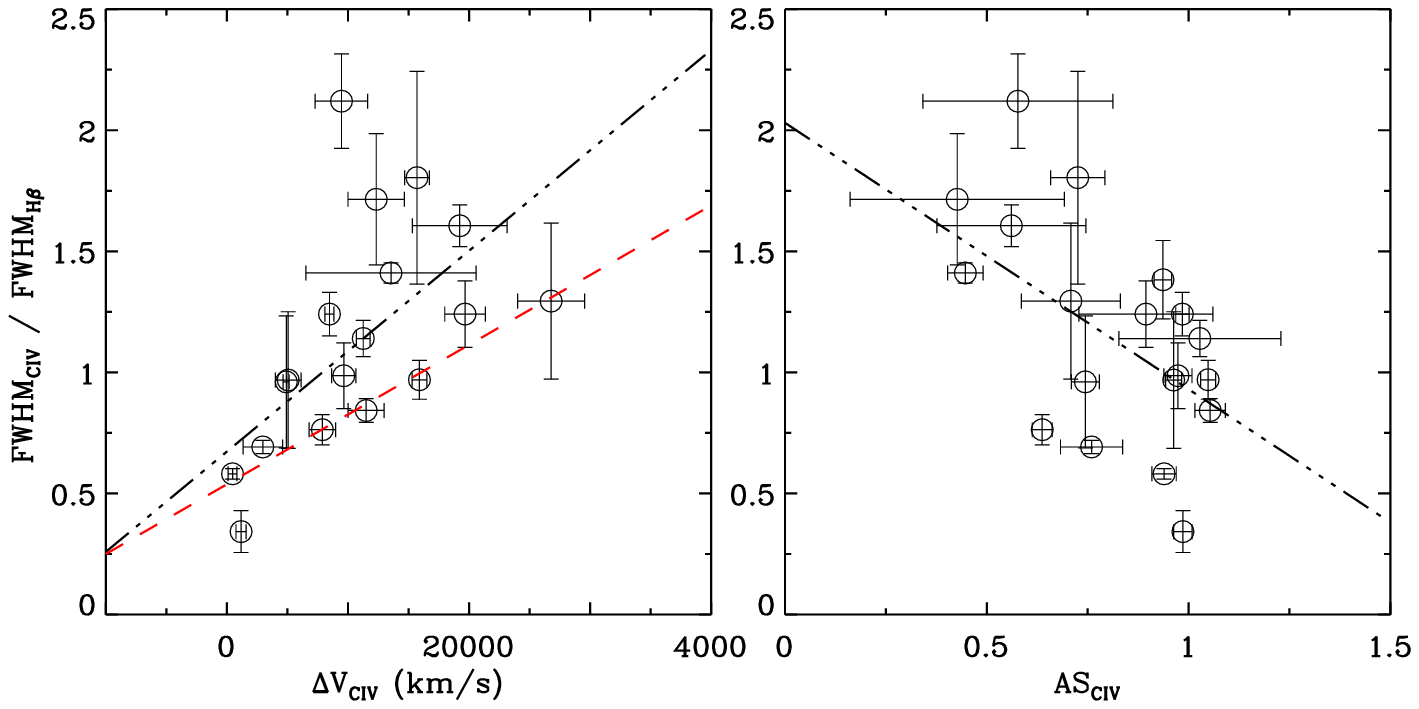}
\caption{Left: Correlation between $\rm FWHM_{CIV} / FWHM_{H\beta}$ and the CIV blueshift for the 18 quasars with positive
$\Delta V_{\rm CIV}$, where the black dot dashed line refers to the best-fitting relation for the 18 quasars and the red dashed line 
represents the best-fitting relation for the 230 high-luminosity, 1.5<z<4.0 quasars shown in Fig. 6 of \cite{Coatman_etal_17}.
Right: Correlation between  $\rm FWHM_{CIV} / FWHM_{H\beta}$ and the CIV asymmetry for the 19 quasars, where the black dot dashed line
refers to the best-fitting relation for the 19 quasars.
\label{fig:fwhmratio_civshift}}
\end{figure*}

The following equation is fitted to the data using the LINMIX$\_$ERR procedure:
\begin{equation} \label{eq:fwhmcor}
\rm FWHM_{\rm CIV} /  FWHM_{\rm H\beta} = \alpha +  \beta\ (\Delta \mit V / \rm 1000\ km\ s^{-1}),
\end{equation}where $\Delta V$ is the CIV blueshift. 
For the 18 quasars with positive $\Delta V$ values, the best-fitting 
results are $\alpha = 0.67\pm0.20$ and $\beta = 0.41\pm0.17$, which is shown as the black line in the left panel 
of Fig. \ref{fig:fwhmratio_civshift}. 
Then the corrected FWHM$_{\rm CIV}$ is calculated as 
FWHM$_{\rm CIV}$ / ($\alpha +  \beta\ (\Delta \mit V / \rm 1000$ km s$^{-1}))$.
The corrected $\rm FWHM_{\rm CIV}$ values based on this calibration are also displayed in the right panel of 
Fig. \ref{fig:l_fwhm_hbciv} as the red symbols.

The red dashed line in Fig. \ref{fig:fwhmratio_civshift} refers to the best-fitting 
relation for the 230 high-luminosity, $1.5<z<4.0$ quasars as shown in Fig. 6 of \cite{Coatman_etal_17} 
with $\alpha = 0.61\pm 0.04$ and $\beta = 0.36\pm 0.03$. The CIV blueshifts in \cite{Coatman_etal_17} are 
defined as $c\ (1549.48 - \lambda_{\rm half}) / 1549.48$, where 1549.48 \AA\ is the rest-frame wavelength 
of the CIV doublet assuming equal contribution from both components. 

Given the anti-correlation between $\rm FWHM_{CIV} / FWHM_{H\beta}$ and the CIV asymmetry, the CIV FWHM can also be
calibrated with the CIV asymmetry to better agree with the H$\beta$ FWHM using the 
LINMIX$\_$ERR procedure:
\begin{equation} \label{eq:fwhmcor_as}
\rm FWHM_{\rm CIV} / FWHM_{\rm H\beta}  = \alpha +  \beta\ AS_{\rm CIV}.
\end{equation}
For the 19 targets, the best-fitting results are $\alpha = 2.03\pm0.52$ and $\beta = -1.1\pm0.62$. The fitting
result is shown as the black line in the right panel of Fig. \ref{fig:fwhmratio_civshift}.

\subsection{Corrections of the CIV-based Virial BH Mass Estimates\label{subsec:results-corrected_bh}}
\begin{figure*}
\plotone{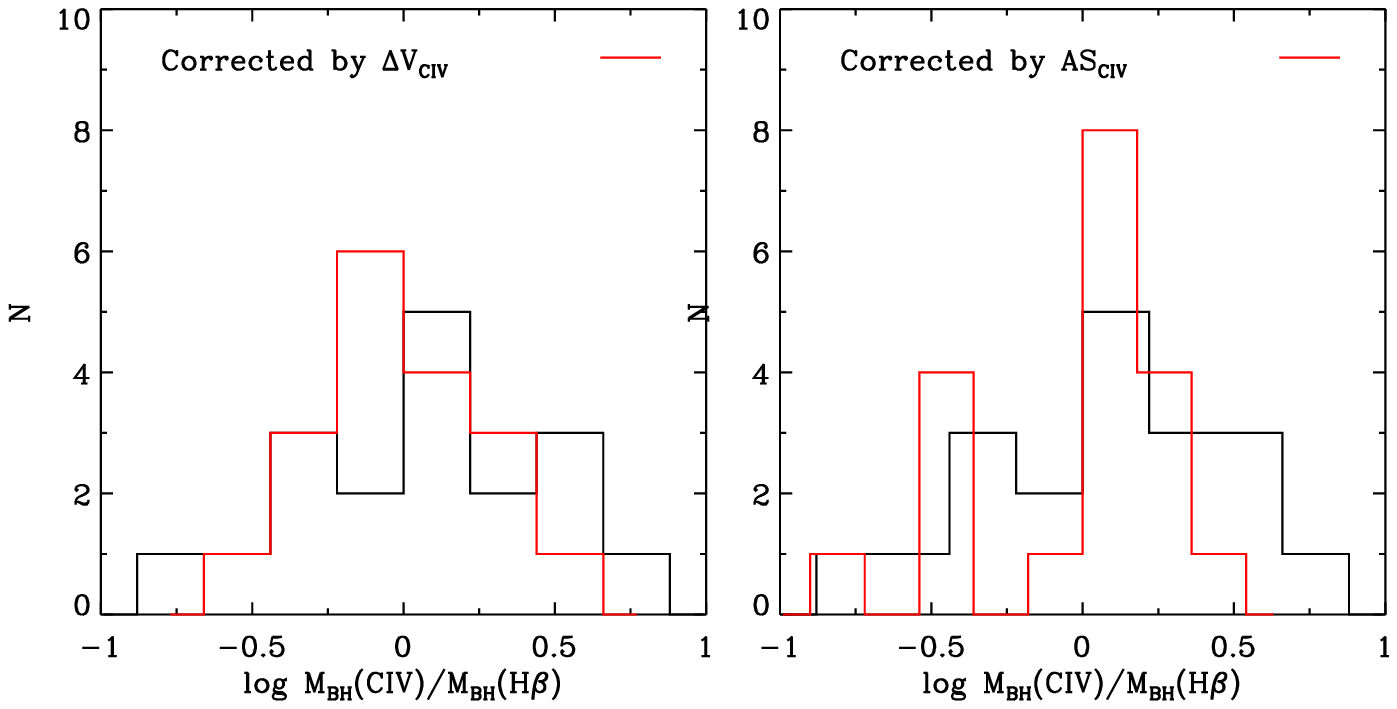}
\caption{Left: For the 18 quasars with positive blueshifts, comparison of the CIV- and H$\beta$-based BH masses before 
(black histogram) and after applying the CIV blueshift-based correction to the CIV FWHMs (red histogram).
Right: For the 19 quasars, comparison of the CIV- and H$\beta$-based BH masses before (black histogram) and after applying
the CIV asymmetry-based correction to the CIV FWHMs (red histogram).}
\label{fig:mciv_corr_mhb}
\end{figure*}

Using the corrected $\rm FWHM_{CIV}$ and the scaling relations in \cite{Vestergaard_Peterson06} for the CIV line, 
we calculate the corrected BH masses $M_{\rm CIV,corr}$. 
The left panel of Fig. \ref{fig:mciv_corr_mhb} compares the CIV- and H$\beta$- based BH masses before and after applying 
the blueshift-based correction to the CIV FWHM for the 18 quasars with positive $\Delta V_{\rm CIV}$. 
The reduction in scatter between the CIV- and H$\beta$-based BH masses can be seen in the reduction in the width of the 
distribution of the mass differences. Before the correction, the median difference between the masses is 0.110 dex 
and the scatter is 0.647 dex. After correcting the CIV FWHM for the non-virial contribution using the $\Delta V_{\rm CIV}$, 
the median difference is reduced to $-0.032$ dex with the scatter as $0.424$ dex. 

The right panel of Fig. \ref{fig:mciv_corr_mhb} compares the CIV- and H$\beta$- based BH masses before and after applying 
the CIV asymmetry-based correction to the CIV FWHM for the 19 quasars. After the correction, the median difference is
$-0.065$ dex with the scatter as $0.643$ dex.  
For the target $J075303.34+423130.8$ with the largest absolute mass difference, the logarithm of the CIV-to-H$\beta$ mass 
ratio before and after the asymmetry-based correction are $-0.845$ and $-0.797$ dex, respectively. Excluding this target, the median differences 
before the asymmetry-based correction is $0.169\pm0.639$ which reduces to $0.123\pm0.359$ after the correction.
Therefore, the correction based on the CIV asymmetry reduces the scatter of $\log M_{\rm BH}({\rm CIV})/ M_{\rm BH}({\rm H}\beta)$ by
less than 0.1 dex and by $\sim0.3$ dex if the target with the largest absolute mass difference is excluded.

\section{Discussion\label{discussion}}
\subsection{Comparison with Previous Works\label{subsec:discussion-comparison}}
Considering the intrinsic error of the SE virial BH mass estimates as $\sim0.4$ dex,
no significant differences between the CIV-based and the 
H$\beta$-based BH masses are found in some previous studies \citep{Vestergaard02, Vestergaard_Peterson06, Greene10, Assef_etal_11}. 
For our high $z$ luminous quasars, the median logarithm of the CIV-to-H$\beta$ 
mass ratios is 0.110 dex but the scatter is 0.647 dex, with 63$\%$ quasars showing mass residuals less than 
0.4 dex. It suggets that the CIV-based BH mass estimates still need to be corrected to reduce the difference with 
the H$\beta$-based BH masses \citep{Denney12, Trakhtenbrot12, Park_etal_13, Runnoe13, Sulentic_etal_17, Coatman_etal_17, Mejia_etal_18, Marziani_etal_19}.

Among the 1350 \AA\ and 5100 \AA\ continuum luminosities, redshift, Eddington ratio, CIV blueshift, CIV asymmetry, 
and the logarithm of the ratio of the UV and optical continuum luminosities, \cite{Assef_etal_11} found only the correlation between 
$\log\ \rm M_{BH}(CIV) / M_{BH} (H\beta)$ and the logarithm of the ratio of continuum luminosities is most significant. They 
suggested that the dispersions in previous comparisons between the CIV- and the H$\beta$-based $M_{\rm BH}$ 
estimates are mainly due to the continuum luminosities rather than any other properties of the lines.

\cite{Shen_Liu12} found that the better correlation between the FWHMs of the CIV and H$\beta$ lines seen 
in \cite{Assef_etal_11} is essentially driven by the objects with lower luminosity. \cite{Marziani_etal_19} suggested that 
the 10 gravitationally lensed quasars in \cite{Assef_etal_11} might have a preferential section of Pop. B quasars 
with FWHM$_{\rm H\beta}>$4000 km s$^{-1}$ and better agreement between H$\beta$ and CIV line widths.

Here, based on an independent high redshift and high luminosity quasar sample, we find a strong correlation between 
$\log\ M_{\rm BH}({\rm CIV}) / M_{\rm BH} ({\rm H}\beta)$ and the CIV blueshift, the CIV asymmetry as well as the CIV 
FWHM, as tabulated in Table \ref{residual}. Different from the result in \cite{Assef_etal_11}, these correlations 
are more significant than that with the logarithm of the ratio of the continuum luminosities.

\begin{figure}
\plotone{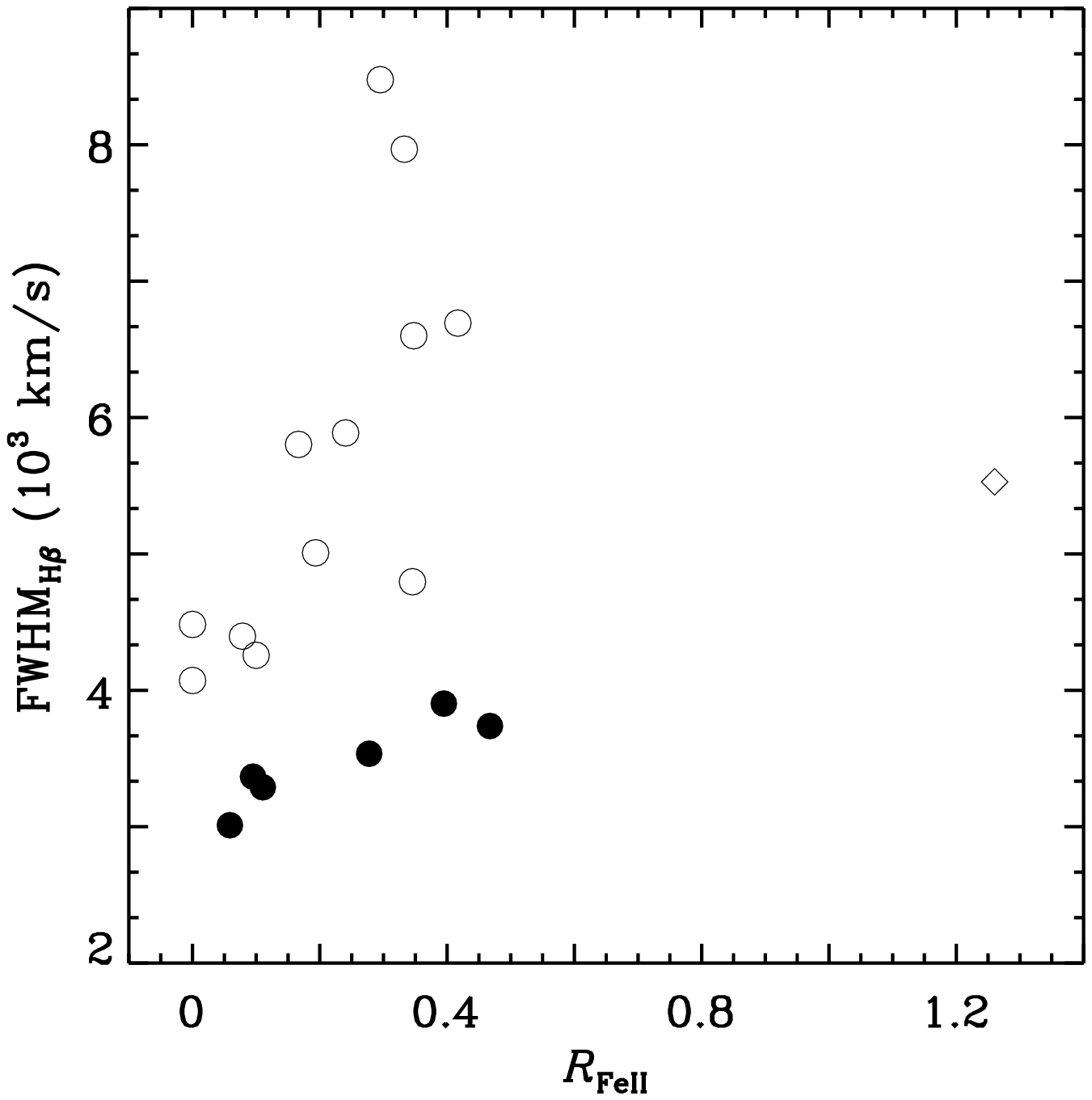}
\caption{Distribution of the quasars in the optical EV1 plane, FWHM$_{\rm H\beta}$ v.s. $R_{\rm FeII}$. The Pop. A1 quasars
with FWHM$_{\rm H\beta}<4000$ km s$^{-1}$ and $R_{\rm FeII}<0.5$ are denoted as filled circles, the Pop. B quasars
with FWHM$_{\rm H\beta}>4000$ km s$^{-1}$ and $R_{\rm FeII}<0.5$ are denoted as open circles and the remaining 1 
quasar is shown as the diamond.
\label{fig:qsoms}}
\end{figure}

The distribution of quasars in the FWHM$_{\rm H\beta}$-$R_{\rm FeII}$ plane in shown in Fig. {\ref{fig:qsoms}}. With 
the classification scheme in \cite{Marziani_etal_10, Marziani_etal_19}, there are 6 Pop. A1 quasars 
(FWHM$_{\rm H\beta}<4000$ km s$^{-1}$, $R_{\rm FeII}<0.5$), 12 Pop. B quasars (FWHM$_{\rm H\beta}>4000$ km s$^{-1}$, 
$R_{\rm FeII}<0.5$) and the remaining one $J081855.77+095848.0$ (FWHM$_{\rm H\beta}>4000$ km s$^{-1}$, $R_{\rm Fe}>0.5$). 
The correlations between FWHM$_{\rm CIV}$ and FWHM$_{\rm H\beta}$ of the quasars in both Pop. A1 and Pop. B are insignificant, 
with $r\sim0.43$ ($p\sim0.40$) and $r\sim0.22$ ($p\sim0.50$), respectively. The correlation between $\log L_{1350}$ and 
$\log L_{5100}$ for the Pop. B quasars is strong with $r\sim 0.76$ ($p\sim0.004$) while there is no correlation between the 
continuum luminosities for the Pop. A1 quasars. Note that the number of Pop. A1 quasars is only 6, which could bias the 
correlation analysis. Nevertheless, the poor relations between the FWHMs in both  the Pop. A1 and Pop. B quasars suggest 
that the differences between the line widths exist and may contribute to the differences between the CIV- and H$\beta$-based BH masses.

From Fig. \ref{fig:mbhratio_all}, for the AGNs with CIV blueshift larger than 1000 km s$^{-1}$, the median CIV-based BH mass is 
roughly overestimated by a factor of $\sim2$ compared to the H$\beta$-based BH masses. It is roughly consistent with the result 
found in \cite{Coatman_etal_16} but with a smaller overestimate value. For quasars with CIV blueshift larger than 2000 km/s, 
the CIV-based BH masses overestimate the H$\alpha$-based BH masses by a factor of $\sim$ 5 \citep{Coatman_etal_16}.

Further using a large sample of 230 high luminosity ($L_{\rm bol} \sim 10^{45.5} - 10^{48}$ erg s$^{-1}$) and $1.5<z<4$ quasars,  
\cite{Coatman_etal_17} corrected the BH masses based on the blueshift-corrected CIV FWHMs, where the blueshifts are calculated 
with $\lambda_{\rm half}$. After the correction, the scatter between the corrected CIV-based and the Balmer line-based BH masses 
decreases from 0.4 dex to 0.24 dex. \cite{Marziani_etal_19} confirmed that the applicability of correcting FWHM${_{\rm CIV}}$ 
using the CIV blueshift and the luminosity for 76 quasars with $L_{\rm bol}\sim10^{44}-10^{48.5}$erg s$^{-1}$ and $0<z<3$.

On correcting the CIV-based BH mass estimates, previous studies have presented other methods based on the parameters 
relatively independent of the redshift estimates, such as reducing the dependence of the BH masses on the emission line 
width \citep{Shen_Liu12, Park_etal_13, Park_etal_17}, the continuum-subtracted peak flux ratio of the ultraviolet emission line 
blend of Si IV$+$O IV relative to CIV \citep{Runnoe13} and the CIV shape \citep{Denney12}. 

\cite{Mejia_etal_18} suggested that these correction methods based on the line peak ratios or blueshifts are of limited 
applicability. It is because most of them depend on correlations that are not driven by an interconnection between the line 
width of CIV and that of the low ionization lines. \cite{Coatman_etal_17} found that the correction based on the CIV blueshifts 
$\Delta V_{\rm CIV}$ yields no systematic bias to correct the BH masses at different blueshifts. It hints that correcting the CIV-based BH 
mass estimates using $\Delta V_{\rm CIV}$ is still valuable. In our work, for the luminous quasars ($L_{\rm bol} > 10^{47.5}$ erg s$^{-1}$),
before the correction, the median difference is 0.110 dex, but the scatter is as large as 0.647 dex. After the correction using 
$\Delta V_{\rm CIV}$, the median difference is $-0.032$ dex and the scatter is reduced from 0.647 dex to 0.424 dex. 
This large scatter indicates that the correction using $\Delta V_{\rm CIV}$ is still not sufficient, and other intrinsic properties have to be taken into account in the corrections of the CIV-based virial BH mass estimates.

Estimates of $\Delta V_{\rm CIV}$ require accurate redshift estimates, which can be obtained with rest-frame optical 
spectra. However, if a rest-frame optical spectrum is available, it is unnecessary to correct the CIV based BH masses 
based on the blueshift. Therefore, a correction using only the information around the CIV line is required, such as the 
AS$_{\rm CIV}$. In our sample, after using the correction with AS$_{\rm CIV}$, the median difference is $-0.065$ dex and 
the scatter decreases to less than 0.1 dex. If the target with the largest absolute mass difference is excluded, the scatter is 
reduced by $\sim$ 0.3 dex. Investigations on whether $\rm AS_{\rm CIV}$ and other properties can be used for the correction of 
$\rm FWHM_{\rm CIV}$, will benefit significantly from a larger sample.

On the other hand, if the systemic redshift can be estimated with an improved method through only the 
rest-frame UV spectra, such as the principal component analysis based redshift estimates, the CIV based BH mass estimates can 
still be corrected using the accurately calculated CIV blueshifts \citep{Allen_etal_2013, Coatman_etal_17, GeXue_etal_19}.

\subsection{Baldwin Effect\label{subsec:discussion-baldwin}}
\begin{figure*}
\plotone{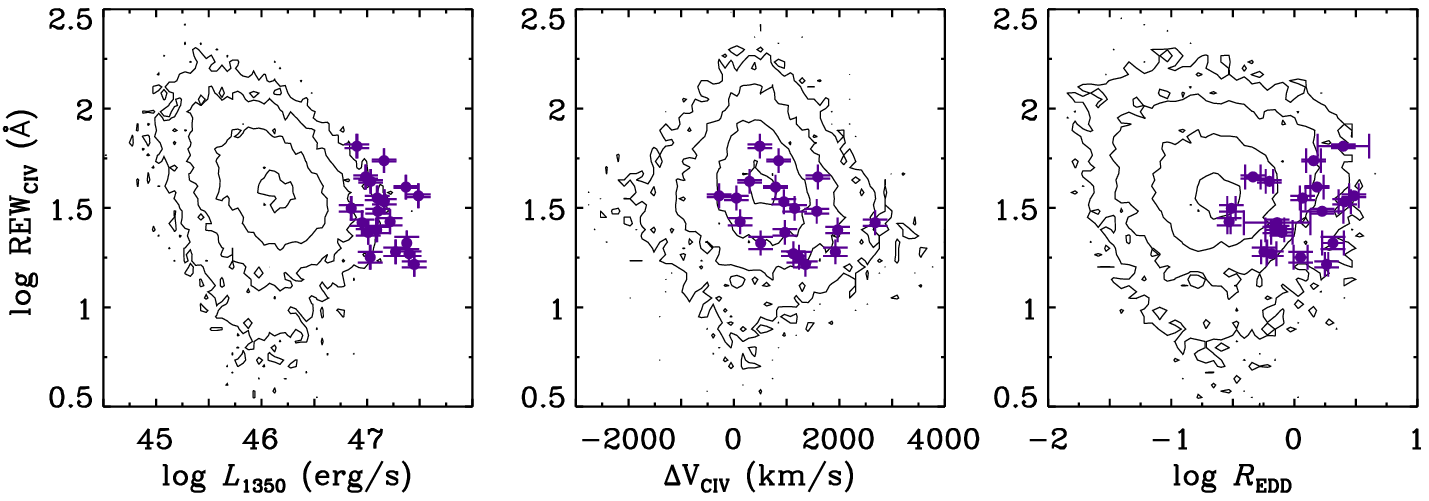}
\caption{Left: Correlation between $\log \rm REW_{\rm CIV}$ and the continuum 
luminosity at 1350 \AA.
Middle: Correlation between $\log \rm REW_{CIV}$ and the CIV blueshift. 
Right: Correlation between $\log \rm REW_{CIV}$ and $\log R_{\rm EDD}$.
In each panel, to show the parameter distribution of our sample with respect to the SDSS DR7 quasars,
analogue data from the DR7 sample of 44424 quasars are superimposed as contours, with the border lines representing the 25, 50, 75, 95 percent contours centered at the maximum probability point.
\label{fig:logewciv_logl1350}}
\end{figure*}

The Baldwin effect is the anti-correlation between the REW values of the CIV line and the 
continuum luminosity at 1450 \AA\ \citep{baldwin_77}. For our sample at z $\sim$ 3.5, a weak 
anti-correlation between REW$_{\rm CIV}$ and the continuum luminosity at 1350 \AA\  can be seen in the 
left panel of Fig. \ref{fig:logewciv_logl1350} ($r\sim-0.31$, $p\sim0.20$). 
For better comparison, in each panel, contours show the distribution of the data from the SDSS DR7 quasar sample of the 
44426 quasars as stated in Section \ref{subsec:results-corrected_fwhm}. The absence of the Baldwin effect may be due to 
the narrow range of the continuum luminosity at 1350 \AA\  in our sample ($10^{46.8} < L_{1350} <10^{47.8}$ erg s$^{-1}$), 
compared to that of the SDSS DR7 sample ($10^{44} < L_{1350} < 10^{47.8}$ erg s$^{-1}$).

As found in previous works \citep{Richards02, Richards_etal_11, Vietri_etal_18}, the REW$_{\rm CIV}$  
values decrease with increasing blueshifts. Using multi-epoch spectra of 362 quasars from the SDSS-RM project, 
\cite{Sun_etal_18} confirmed that the extremely blueshifted quasars generally have smaller REW$_{\rm CIV}$, while 
the reverse is not true. With the CIV blueshift range narrower than that of the quasar sample in \cite{Sun_etal_18}, 
we find that the REW$_{\rm CIV}$  values moderately anti-correlates with the CIV blueshifts with $r\sim-0.45$ ($p\sim0.06$). 
The result is shown in the middle panel of Fig. \ref{fig:logewciv_logl1350}.

Some previous works proposed a modified Baldwin effect, relating REW$_{\rm CIV}$ and $\log R_{\rm EDD}$ 
\citep{Shemmer_Lieber_15, GeXue_etal_16}. 
For our sample, the bolometric luminosity $L_{\rm bol}$ and the Eddington ratio $R_{\rm EDD}$ are directly taken from 
\cite{Zuo_etal_15}. The $L_{\rm bol}$ values are estimated with $L_{5100}$ using the 
bolometric correction factor of 9.26 from the composite spectral energy distribution (SED) \citep{Richards_etal_06}, 
which presented that the uncertainty of the bolometric luminosity can be as much as 0.3 dex under the assumption 
of a single mean SED. With the bolometric luminosities and the H$\beta$-based BH masses, the $R_{\rm EDD}$ values
are estimated as $L_{\rm bol} / L_{\rm EDD}$. The errors of $L_{\rm bol}$ and $R_{\rm EDD}$ listed in Table 2 of 
\cite{Zuo_etal_15} only account for the statistical uncertainties estimated using the Monte Carlo approach. Including
the uncertainty in the bolometric correction ($\sim0.3$ dex) and the intrinsic uncertainty of the virial BH mass estimates
($\sim0.4$ dex), the error propagation would yield the errors of $R_{\rm EDD}$ larger than $\sim0.5$ dex.

No similar anti-correlation between the two parameters is seen in our sample, 
as shown in the right panel of Fig.  \ref{fig:logewciv_logl1350}. We argue that it is probably due to the narrow range 
of $\log R_{\rm EDD}$ in our sample; the range of $\log R_{\rm EDD}$ in our sample is [-0.52, 0.49] compared to 
[-2, 0.5] in \cite{GeXue_etal_16} and [-2, 0.6] in the SDSS DR7 quasar sample \citep{Shen11}.
A larger sample with a wider $\log R_{\rm EDD}$ range is needed to better understand the modified Baldwin effect. 

\subsection{Radio Properties\label{subsec:discussion-radio}}
To study the differences between quasars with small and large CIV blueshifts, we divide the 18 quasars with positive 
$\Delta V_{\rm CIV}$ into 2 subsamples (subsample I and II) according to $\Delta V_{\rm CIV}$ separated at 1000 km/s; 
subsample II (9 quasars) has higher CIV blueshifts than that of subsample I (9 quasars). 

Some previous studies showed that radio-loud quasars are strongly biased to have lower mean CIV blueshifts than the 
radio-quiet quasars \citep{Marziani_etal_96, Richards_etal_11, Sulentic_etal_07}. 
As shown in Table \ref{table:continuum-line-params}, among the 9 quasars
in subsample I, 1 quasar is not in the FIRST footprint, 6 quasars are not detected in the FIRST and 2 quasars are radio detected with
the radio loudness $R$ as 5.5 and 2645, respectively.
Among the 9 quasars in subsample II, 1 is not in the FIRST footprint, 7 are not detected in the FIRST and 1 is radio-detected
with $R\sim14.0$. 

With a limited sample, our results show that the fraction of radio-detected quasars in subsample I is larger than that in subsample II and
the loudness of the radio-loud quasar in subsample I is larger than that in subsample II. However, whether it 
suggests that a higher blueshift sample is more radio quiet, still requires further verification based on a larger sample 
of quasars.

Using 130 low z AGNs, \cite{Sulentic_etal_07} found that sources with CIV blueshifts strongly favor radio-quiet 
Pop. A quasars with FWHM$_{\rm H\beta}\le4000$ km s$^{-1}$ and the CIV blueshift is not observed in most Pop. B sources. 
For $0.1<z<3.1$ and $47.4<\log L_{\rm bol}<48.4$ erg s$^{-1}$ AGNs, \cite{Sulentic_etal_17} confirmed the preference 
of CIV blueshifts in Pop. A quasars and found that many Pop. B quasars show significant CIV blueshifts.
Using the WISSH quasar sample with $L_{\rm bol}>10^{47.3}$ erg s$^{-1}$, \cite{Vietri_etal_18} also found 
large CIV blueshifts in sources with FWHM$_{\rm H\beta}>4000$ km s$^{-1}$. 

Among 6 Pop. A1 quasars, only 2 quasars (33.3\% of the Pop. A1 quasars) are in subsample II. Among 12 Pop. B quasars, 
6 quasars (50\% of the Pop. B quasars) are in subsample II with CIV blueshifts greater than 1000 km s$^{-1}$, 
extending the detection of significant CIV blueshifts in luminous Pop. B quasars.

\subsection{Broad Emission Line Region Models\label{subsec:discussion-BELR_models}}
\begin{figure}
\plotone{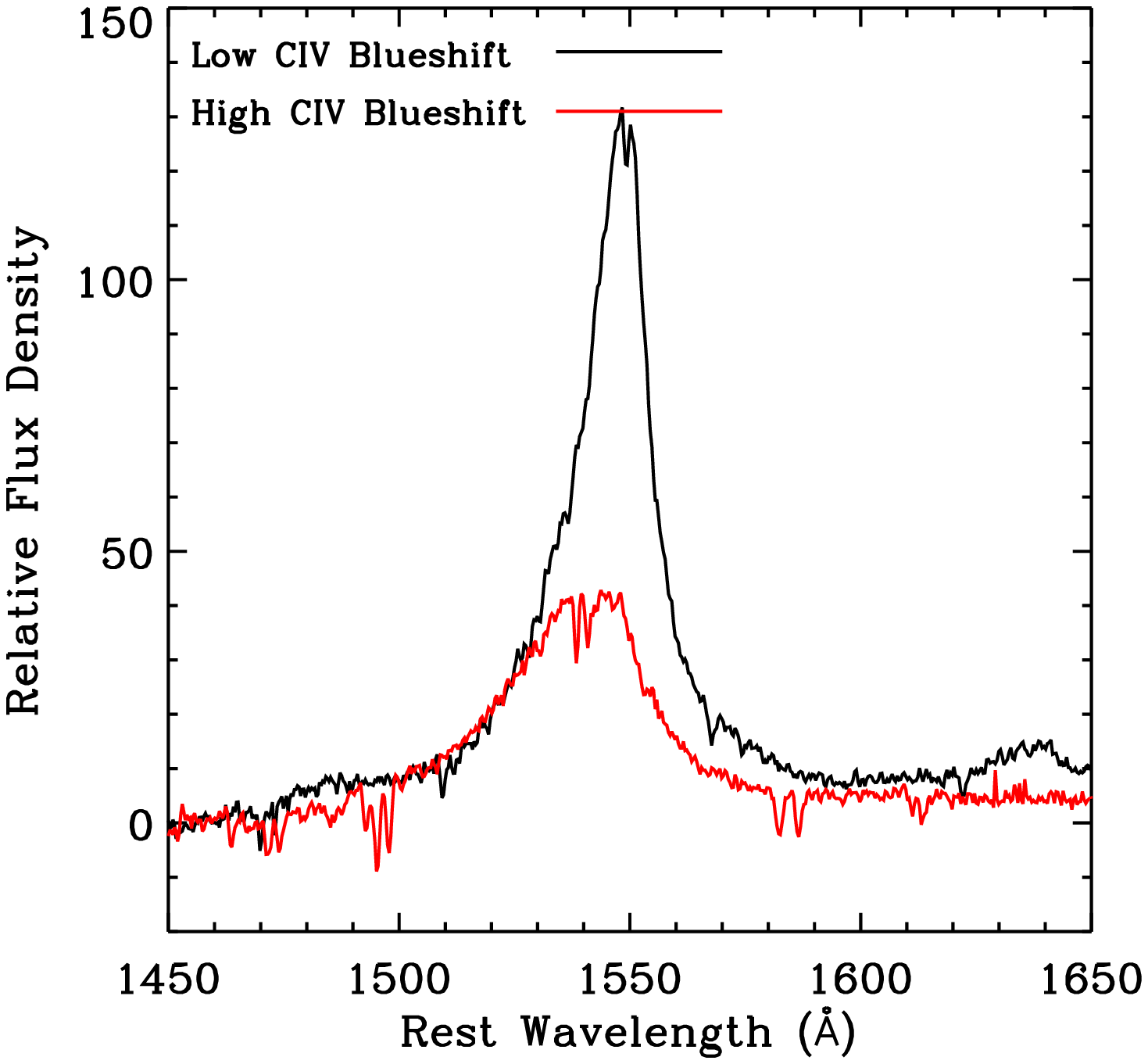}
\caption{CIV emission lines of the composite spectra for subsamples with high and low CIV blueshifts. 
Black and red lines represent the composite spectra with low ($<$ 1000 km/s) and high ($>$ 1000 km/s) 
$\Delta V_{\rm CIV}$ values, respectively.
\label{fig:stackspec_line}}
\end{figure}

As we are more interested in the emission line properties but not the continuum of the quasar spectra, we create 2 
composite spectra based on the rest-frame spectra of the two subsamples using the arithmetic mean instead of the 
geometric mean at each wavelength pixel. After subtracting the best-fitting pseudo-continuum from the stacked spectra, 
we investigate their CIV emission line profiles. As shown in Fig. \ref{fig:stackspec_line}, the blue wings are similar, 
while the red wing of the CIV line profile with larger blueshift is lower. It appears that the shift of the CIV emission 
line is not due to the blueshift of the whole profile but to the suppression of the red wing. 

\cite{Richards02} suggested that the CIV blueshifts were related to the quasar orientation, either external or 
internal. In a spherically symmetric cloud model \citep{Elvis00,Richards02}, assuming the outflowing clouds are 
isotropically distributed, subsample I may represent a more face-on configuration of the accretion disk and subsample 
II represents a more edge-on configuration \citep{Richards02, Leighly04, Coatman_etal_17,  Coatman_etal_16}. The obscuration 
of the optical-thick disk in subsample II would reduce the flux of the red wing of CIV. While for a disk wind-type model 
\citep{Murray_Chiang_98}, the blueshift and shape could be caused by both the orientation and a change in the opening 
angle of the disk wind.

However, \cite{Leighly04} and \cite{Richards_etal_11} hinted that the CIV blueshifts are not likely due to the external 
orientation. \cite{Leighly04} mentioned that the blueshifted high-ionization lines may come from a wind that is moving 
toward us, with the receding side obscured by the optically thick accretion disk. Moreover, 
some studies suggested that the blueshift implied the contribution from the outflowing wind component 
\citep{Richards_etal_11, Bachev_etal_04, Marziani_etal_10, Vietri_etal_18, Marziani_etal_19}. 

In that way, the differences of the CIV properties between subsamples I and II are due to differences in the accretion disk wind, 
where the relative contribution of the wind component to the disk contribution in subsample II is larger. Furthermore, 
the difference of the disk wind is probably due to the difference of the spectral energy distribution, ultimately 
determined by the intrinsic quasar properties, such as the Eddington ratio 
\citep{Richards_etal_11, Sulentic_etal_17, Vietri_etal_18, Sun_etal_18, Marziani_etal_19}.  

On the other hand, we compared the REW$_{\rm [OIII]}$ distribution of the two subsamples. We find that depending on 
REW$_{\rm [OIII]}>5$ or $<5$  \AA \citep{Vietri_etal_18}, subsample I can be divided in 6 [OIII] and 3 weak [OIII] quasars; 
while subsample II contains 1 [OIII] quasar and 8 weak [OIII] quasars. The higher ratio of [OIII] 
quasars in subsample I with lower blueshifts suggests that the [OIII] quasars seem to exhibit lower blueshifts than the 
weak [OIII] quasars. This is qualitatively consistent with the result found in \cite{Vietri_etal_18}, where among 18 quasars 
with $2<z<4$ and $L_{\rm bol}>10^{47.3}$ erg s$^{-1}$, the 6 [OIII] quasars show lower blueshifts than the other 12 weak [OIII] quasars. 

\cite{Shen_Ho14} suggested that the dispersion of the FWHM$_{\rm H\beta}$ at fixed $R_{\rm FeII}$ in the optical EV1 plane 
is largely an orientation effect and found no tendency of the REW$_{\rm [OIII]}$ values with 
FWHM$_{\rm H\beta}$ at fixed $R_{\rm FeII}$. That indicates that the REW$_{\rm [OIII]}$ distribution can not be 
solely explained by the orientation effect.

Since the 18 quasars of our sample exhibit $R_{\rm FeII}$ values ranging from 0 to 0.47, as shown in 
Fig. \ref{fig:qsoms}, it is reasonable to assume them as quasars with almost fixed $R_{\rm FeII}$ and to take the FWHM$_{\rm H\beta}$ 
as an orientation indicator. No significant correlations between FWHM$_{\rm H\beta}$ and the CIV blueshifts as well as 
REW$_{\rm [OIII]}$ are found, with $r\sim0.25$ ($p\sim0.32$) and $r\sim0.09$ ($p\sim0.73$), respectively. 
It indicates that neither the CIV blueshifts nor the distribution of REW$_{\rm [OIII]}$ would be likely explained by the 
orientation effect alone.

Though our results can not help constrain the details of the Broad Emission Line Region models, the 
strongly blueshifted CIV line profile suggests that it is at least partly contributed by the emission in the wind. The 
possibility of an orientation effect can not be neglected, but the orientation effect alone is unlikely to explain 
either the CIV blueshift or the REW$_{\rm [OIII]}$ distribution. It is possible that they are caused by the 
combination of the orientation effect and the intrinsic quasar properties, such as the Eddington ratio.

\subsection{Is the CIV blueshift driven by the Eddington Ratio?}
\begin{figure}
\plotone{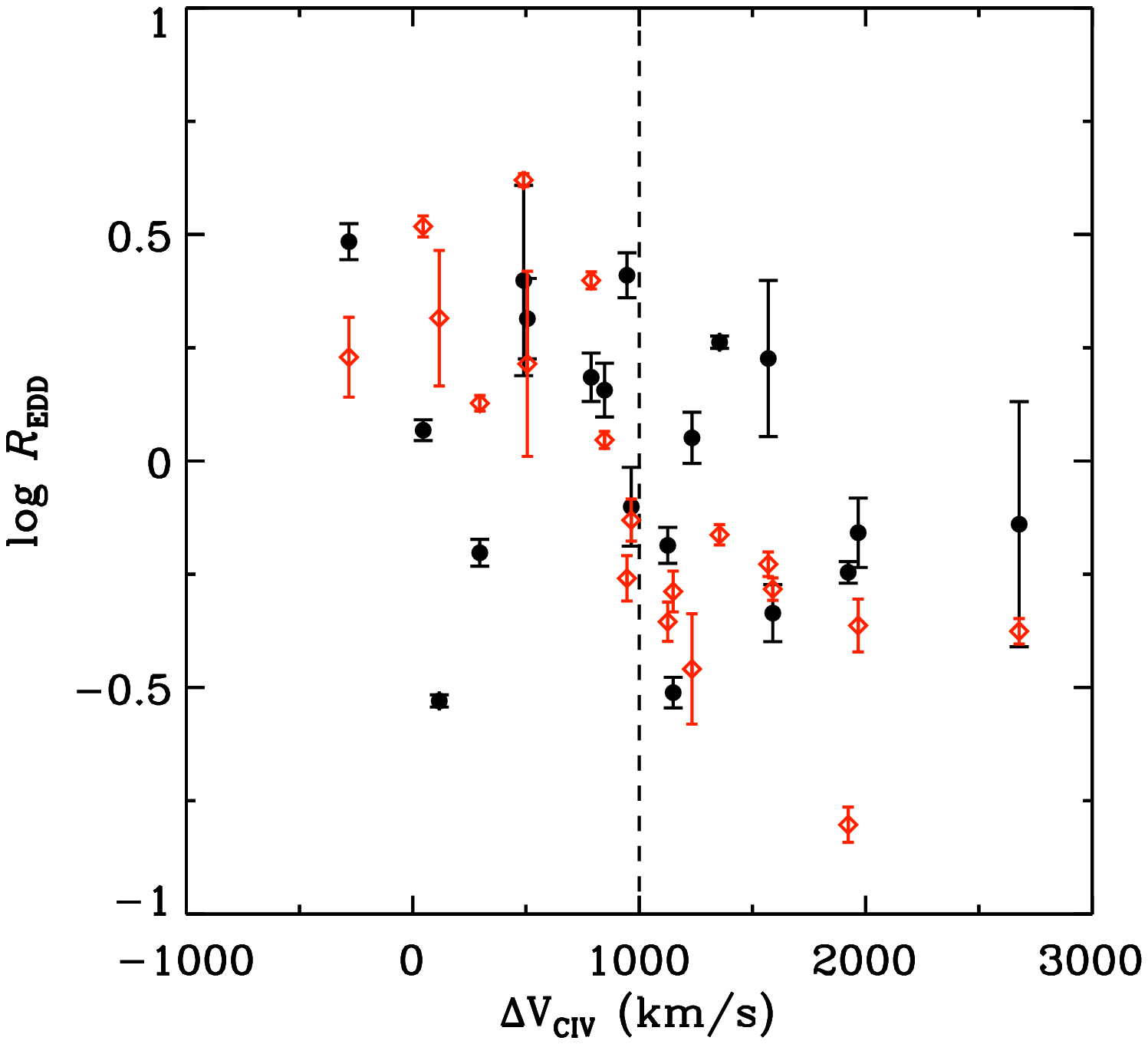}
\caption{Eddington ratio v.s. the CIV blueshift, where the black 
and red symbols refer to the H$\beta$-based $\log R_{\rm EDD}$ and the CIV-based $\log R_{\rm EDD}$ values, respectively. The 
quasars in our sample are accreting at $R_{\rm EDD} > 0.3$ and covers a large range of blueshifts.
\label{logcivshift_logedd}}
\end{figure}

As shown in Fig. \ref{logcivshift_logedd}, adopting $R_{\rm EDD}$ estimated from the H$\beta$-based BH masses, all 
the quasars in our sample are accreting at $R_{\rm EDD}>0.30$ and cover a considerable range of blueshifts. It suggests 
that not all quasars with high Eddington ratios show large blueshifts. This is consistent with previous works 
\citep{Baskin05, Coatman_etal_16, Sun_etal_18}, where quasars with large CIV blueshifts tend to accrete at around 
Eddington limits, while the converse is not true. 

The Eddington ratio is thought to be the underlying driver behind the MS \citep{Boroson_Green92, Sulentic_etal_00, 
Marziani_etal_01, Bachev_etal_04, Marziani_etal_10, Shen_Ho14, Sulentic_etal_17, Marziani_etal_19}. 
As the $R_{\rm EDD}$ value increases, the radiation pressure plays an increasingly important role in accelerating clouds which could
produce blueshifted CIV profile \citep{Marziani_etal_10}. Previous studies found that there is a correlation between the 
CIV line blueshifts and the $R_{\rm EDD}$ values \citep{Marziani_etal_10, Coatman_etal_16, Sun_etal_18, Vietri_etal_18}.

No similar correlation between the CIV blueshifts and $R_{\rm EDD}$  is seen in our sample, with $r\sim-0.2$ ($p\sim0.4$) . 
The median $\log R_{\rm EDD}$ of the quasars in subsample I is $0.16\pm0.42$, while that of 
the quasars in subsample II is  $-0.16\pm0.30$. The scatter is estimated as the inner 50 percentile of the distribution.
From the difference of the median $\log R_{\rm EDD}$, it seems that the quasars of subsample I accrete at higher 
Eddington ratio than the quasars of subsample II. However, this may not be representative of the 
intrinsic relationship, as  both the errors of $R_{\rm EDD}$ ($\sim0.5$ dex) and the scatters of the 
median $\log R_{\rm EDD}$ are large compared to the difference of the median $\log R_{\rm EDD}$ 
between subsample I and II. 
If the $R_{\rm EDD}$ values are estimated from the CIV-based BH masses, 
as shown in red, an anti-correlation between the CIV line blueshifts and the $R_{\rm EDD}$ is shown, with 
$r\sim-0.86$ ($p<0.01$). This anti-correlation is mainly due to the positive correlation between the CIV line blueshifts 
and the CIV-based BH masses, as shown in Section \ref{subsec:results-mass_ratio}. 

In the optical EV1 plane, the average $R_{\rm EDD}$ value increases as the $R_{\rm FeII}$ increases
\citep{Shen_Ho14, Marziani_etal_19}. Our sample including only the Pop. A1 and Pop. B quasars with $R_{\rm FeII}<0.5$ 
covers the left part in the optical EV1 plane. Thus we argue that the inconsistent correlation 
between the CIV blueshifts and $R_{\rm EDD}$ in our sample is mainly due to the limited sample size 
and the range of the $R_{\rm EDD}$, which should be verified with a larger sample in the future.

\section{CONCLUSION}

Using a $z\sim3.5$ luminous quasar sample ($10^{47.5} < L_{\rm bol} < 10^{48.3}$ erg s$^{-1}$) with observed-frame 
optical and NIR spectroscopy, we have investigated the reliability of estimating BH masses based on the CIV emission line
and the possible corrections based on the CIV blueshift and asymmetry. 
We also study the CIV emission line properties, in terms of the blueshift and possible physical mechanisms therein. 
The conclusions are summarized as follows:

1. The logarithm of the CIV-to-H$\beta$ BH mass ratios are between $-0.85$ dex and $0.67$ dex, with 
a median value as 0.110 dex and a scatter as 0.647 dex. 63\% quasars in our sample show 
$\log M_{\rm BH}({\rm CIV}) / M_{\rm BH}({\rm H}\beta)$ within the range of 0.4 dex. It suggests that the CIV-based 
BH mass estimates still need to be corrected to better match the H$\beta$-based BH masses.

2. The logarithm of the CIV-to-H$\beta$ BH mass ratios correlate with the CIV FWHMs, the CIV blueshifts and the 
asymmetries. Corrections using the CIV blueshift and the asymmetry reduce the scatter of the logarithm of the mass ratio 
by $\sim$0.2 dex and $\sim$0.04 dex, respectively. Excluding the target with the largest CIV-to-H$\beta$ mass difference, 
the scatter is reduced by $\sim0.3$ dex after the correction using the CIV asymmetry. 

3. The Baldwin effect between REW$_{\rm CIV}$ and $\log L_{1350}$ is not significant in our sample, likely due to the 
limited luminosity range of our sample. The moderate anti-correlation between REW$_{\rm CIV}$ and the CIV blueshifts is seen. No similar anti-correlation between REW$_{\rm CIV}$ and $R_{\rm EDD}$ (the modified Baldwin 
effect) is seen, which may be mainly due to the narrow range of $\log R_{\rm EDD}$ of our sample.

4.With limited number of quasars, we find that in subsample I with positive CIV blueshifts$<$1000 km s$^{-1}$ among the 8 
quasars in the FIRST footprint, 2 quasars are radio-detected including 1 radio-loud quasar with $R\sim2645$
and 1 radio-quiet quasar with $R\sim5.5$, while in subsample II among the 8 quasars in the FIRST footprint, 
1 quasar is radio-detected in the FIRST with $R\sim14.0$. Whether it suggests that a higher blueshift sample is more 
radio quiet, still needs to be verified using a larger quasar sample in the future. 

5. Among the 19 quasars, 6 quasars are the Pop. A1 and 12 are the Pop. B quasars. 50$\%$ of the Pop. B quasars exhibit 
the CIV blueshifts larger than 1000 km s$^{-1}$, extending the detection of significant CIV blueshifts in luminous Pop. B quasars.

6. After comparing the CIV emission line profiles of the composite spectra of subsamples I and II, we find a lack of flux in the 
red wing for the CIV composite emission line of subsample II with larger blueshifts. The ratio of [OIII] quasars with 
REW$_{\rm [OIII]} > 5$ \AA\ in subsample I is higher by 55.6$\%$ than that in subsample II, suggesting that the [OIII] quasars 
seem to exhibit lower blueshifts than the weak [OIII] quasars with REW$_{\rm [OIII]} < 5$ \AA.

7. Considering that our sample mainly covers a narrow range of $R_{\rm FeII}<0.5$, we take the FWHM$_{\rm H\beta}$ as an indicator 
of the orientation. No significant correlation between FWHM$_{\rm H\beta}$ and the CIV blueshifts or REW$_{\rm [OIII]}$ is found, 
indicating that the orientation effect alone can not explain either the difference of the CIV profile for quasars with 
different blueshifts or the distribution of REW$_{\rm [OIII]}$. It is more likely that they are caused by the orientation 
effect and the intrinsic quasar properties, such as the Eddington ratio.

8. Quasars in our sample accrete at high Eddington ratios with $R_{\rm EDD} > 0.3$ and show a wide range of CIV blueshifts, with 18/19 of the quasars showing CIV blueshifts (with the median value of 1126 km s$^{-1}$) and 14/19 of the quasars showing CIV blueshifts larger than 500 km s$^{-1}$.\\

\begin{deluxetable}{ccccccc}
  \tabletypesize{\footnotesize}
  \tablecaption{Spectra Information from the SDSS DR7 and DR7$+$ \label{table:spectra_dr7and14}}
  \tablewidth{0pt}
  \tablehead{\colhead{Name (SDSS)} & \colhead{$N_{\rm obs}$} & \colhead{S/N} & \colhead{Note} \\
  \colhead{(1)} & \colhead{(2)} & \colhead{(3)} & \colhead{(4)}}\\
\startdata
 J011521.20+152453.3 & 2  & 23.61 & DR7, highest S/N \\  \\
 J014214.75+002324.2 & 1  & 20.37 & DR7              \\  \\
 J015741.57-010629.6 & 3  & 19.22 & DR7              \\  \\
                     &    & 22.61 & DR7$+$, highest S/N\\  \\
 J025021.76-075749.9 & 2  & 20.62 & DR7, highest S/N        \\  \\ 
 J025905.63+001121.9 & 6  & 23.79 & DR7              \\  \\
                     &    & 37.03 & DR7$+$, highest S/N\\  \\
 J030341.04-002321.9 & 4  & 25.87 & DR7              \\  \\
                     &    & 27.99 & DR7$+$, highest S/N\\  \\
 J030449.85-000813.4 & 7  & 29.62 & DR7, highest S/N \\  \\
 J075303.34+423130.8 & 4  & 32.94 & DR7              \\  \\
                     &    & 40.69 & DR7$+$, highest S/N\\  \\
 J080430.56+542041.1 & 2  & 22.62 & DR7, highest S/N \\  \\
 J080819.69+373047.3 & 4  & 12.55 & DR7              \\  \\
                     &    & 19.35 & DR7$+$, highest S/N\\  \\
 J080956.02+502000.9 & 3  &  18.8 & DR7              \\  \\
                     &    & 27.56 & DR7$+$, highest S/N\\  \\
 J081855.77+095848.0 & 1  & 20.67 & DR7              \\  \\
 J090033.50+421547.0 & 2  & 40.85 & DR7              \\  \\
                     &    & 47.48 & DR7$+$, highest S/N\\  \\
 J094202.04+042244.5 & 2  & 28.96 & DR7              \\  \\
                     &    & 41.07 & DR7$+$, highest S/N\\  \\
 J115954.33+201921.1 & 1  & 39.94 & DR7              \\  \\
 J173352.23+540030.4 & 4  &  35.5 & DR7              \\  \\
                     &    & 36.11 & DR7$+$, highest S/N\\  \\
 J213023.61+122252.0 & 1  & 21.24 & DR7              \\  \\
 J224956.08+000218.0 & 2  & 14.56 & DR7              \\  \\
                     &    & 24.95 & DR7$+$, highest S/N\\  \\
 J230301.45-093930.7 & 1  & 26.33 & DR7              \\  \\
\enddata
\tablecomments{
  Col.(1) Name of the quasars.
  Col.(2) Number of spectra for this quasar in the SDSS DR14 database.
  Col.(3) Average S/N per pixel of this spectrum.
  Col.(4)`DR7' means that the row refers to the information of the spectrum used in the SDSS DR7 catalog (Shen et al. 2011). `DR7$+$' means that the row refers to the information of the spectrum with the highest S/N per pixel after the SDSS DR7.}
\end{deluxetable}

\begin{deluxetable*}{ccccccc}
\tablecaption{The fitting details of the CIV emission line \label{table:civdetails}}
\tablewidth{0pt}
\tablehead{
\colhead{Name (SDSS)} &  
\colhead{$\Delta V_{0}$ | REW$_{0}$ | FWHM$_{0}$} &
\colhead{$\Delta V_{1}$ | REW$_{1}$ | FWHM$_{1}$} &
\colhead{$\Delta V_{2}$ | REW$_{2}$ | FWHM$_{2}$} &
\colhead{BC indexes} &
\colhead{$\Delta V$  | REW | FWHM} & 
\colhead{Model}  \\
\colhead{} & 
\colhead{(km s$^{-1}$ | \AA | km s$^{-1}$)} & 
\colhead{(km s$^{-1}$ | \AA | km s$^{-1}$)} & 
\colhead{(km s$^{-1}$ | \AA | km s$^{-1}$)} & 
\colhead{} &
\colhead{(km s$^{-1}$ | \AA | km s$^{-1}$)} & 
\colhead{}  \\ 
\colhead{(1)} & \colhead{(2)} & \colhead{(3)} & \colhead{(4)} & \colhead{(5)} & \colhead{(6)} & 
\colhead{(7)}   
} 
\startdata
 J011521.20+152453.3        & $4324$ / 11 /  6362 & $ 1852$ /  9 /  4227 & $3112$ /  7 / 22831 & 0,1,2 & $2433$ / 27 / 6483 &  A \\ \\ 
 J014214.75+002324.2        & $ 919$ /  7 /  3261 & $ 1354$ / 16 / 11486 & $ 795$ /  0 /  6496 & 0,1,2 & $ 962$ / 24 / 4730 &  A \\ \\ 
 {\bf J015741.57-010629.6}  & $1425$ / 26 / 12906 & $ 1938$ / 19 /  4952 & $ 781$ /  4 /  1600 & 0,1   & $1918$ / 45 / 6489 &  B \\ \\ 
 J025021.76-075749.9        & $-125$ /  3 /  2722 & $ 2128$ / 15 /  8950 & $-227$ /  1 /   853 & 0,1   & $ 140$ / 18 / 6983 &  B \\ \\ 
 {\bf J025905.63+001121.9}  & $ 829$ / 28 / 13567 & $  995$ / 26 /  4486 & $ -92$ / 10 /  1600 & 0,1   & $ 957$ / 55 / 5562 &  B \\ \\ 
 J030341.04-002321.9        & $ 196$ /  3 /  2310 & $ 1537$ / 17 /  6233 & $ 905$ / 14 / 15617 & 0,1,2 & $ 511$ / 34 / 6383 &  A \\ \\ 
{\bf J030449.85-000813.4}   & $ 933$ / 13 / 11092 & $  422$ /  8 /  2563 & $-114$ /  4 /  1285 & 0,1   & $ 421$ / 21 / 3261 &  B \\ \\ 
 J075303.34+423130.8        & $ 421$ / 18 /  8671 & $  -10$ /  9 /  2114 & $ 194$ /  0 /  1600 & 0,1   & $  22$ / 27 / 2903 &  B \\ \\ 
 J080430.56+542041.1        & $3905$ / 15 /  9893 & $-4716$ /  5 / 15309 & $ 984$ / 10 /  4612 & 0,1,2 & $1258$ / 30 / 7041 &  A \\ \\ 
 J080819.69+373047.3        & $1873$ / 20 /  5937 & $-4673$ / 11 / 21070 & $ 421$ /  0 /   984 & 0,1   & $1755$ / 31 / 6714 &  B \\ \\ 
 {\bf J080956.02+502000.9}  & $ -34$ /  9 /  2362 & $ -437$ / 17 / 12375 & $ 970$ / 17 /  4872 & 0,1,2 & $ 147$ / 43 / 4016 &  A \\ \\ 
 J081855.77+095848.0        & $3902$ / 13 / 14787 & $ 1821$ / 11 /  5206 & $ 197$ /  0 /  5631 & 0,1,2 & $1954$ / 25 / 6860 &  A \\ \\ 
 {\bf J090033.50+421547.0}  & $-347$ / 11 /  3319 & $    8$ / 26 / 11593 & $-988$ /  1 /  1600 & 0,1   & $-360$ / 36 / 4888 &  B \\ \\ 
 {\bf J094202.04+042244.5}  & $1181$ / 16 /  4595 & $  274$ /  7 /  1761 & $1686$ / 17 / 13385 & 0,1,2 & $ 372$ / 40 / 3356 &  A \\ \\ 
 J115954.33+201921.1        & $1758$ /  7 / 27583 & $ 1385$ / 12 /  6504 & $-251$ /  0 /  1410 & 0,1,2 & $1447$ / 19 / 7525 &  A \\ \\ 
 {\bf J173352.23+540030.4}  & $1463$ /  5 /  4095 & $ 3385$ /  9 / 12457 & $-128$ /  2 /  2165 & 0,1,2 & $ 254$ / 16 / 5273 &  A \\ \\ 
 J213023.61+122252.0        & $4591$ / 10 / 29638 & $  483$ / 17 /  5852 & $ -23$ /  8 /  1647 & 0,1,2 & $  16$ / 35 / 2474 &  A \\ \\ 
 J224956.08+000218.0        & $1264$ / 22 /  4067 & $   57$ / 18 /  2070 & $ 467$ / 24 / 14643 & 0,1,2 & $ 226$ / 65 / 3160 &  A \\ \\ 
 J230301.45-093930.7        & $1424$ /  4 /  4175 & $ -443$ /  0 /  4973 & $3675$ / 15 / 13763 & 0,1,2 & $1685$ / 19 / 9453 &  A \\ \\
\enddata
\tablecomments{
Col.(1) Name of the quasars. The name in bold refer to the quasars, the [OIII] doublets of which were fitted with two pairs of Gaussians.
Col.(2) CIV emission-line velocity shift $\Delta V^{\rm peak}$ with respect to [OIII] $\lambda$5007, REW and FWHM of the 1$^{\rm st}$ Gaussian for the CIV line fitting.
Col.(3) $\Delta V^{\rm peak}$, REW and FWHM of the 2$^{\rm nd}$ Gaussian for the CIV line fitting.
Col.(4) $\Delta V^{\rm peak}$, REW and FWHM of the 3$^{\rm rd}$ Gaussian for the CIV line fitting.
Col.(5) The indexes of Gaussians for the CIV BC.
Col.(6) $\Delta V^{\rm peak}$, REW and FWHM of the CIV BC.
Col.(7) Fitting model of the CIV emission line. In Model A, 3 Gaussians are fitted to the CIV BC and the contribution of the CIV NC is not considered, while in Model B, 2/1 Gaussians are fitted to the CIV BC/NC.
}
\end{deluxetable*}

\begin{deluxetable*}{cccccccccc}
\tablecaption{The continuum and emission line parameters \label{table:continuum-line-params}}
\tablewidth{0pt}
\tablehead{
\colhead{Name (SDSS)}   &  \colhead{$z$} & \colhead{$\log L_{1350}$}  & \colhead{FWHM$_{\rm CIV}$} &
\colhead{$\log \rm M^{\rm CIV}_{BH} (VP06)$} & \colhead{REW$_{\rm CIV}$} & 
\colhead{$\Delta V^{\rm peak}_{\rm CIV}$ | $\Delta V^{\rm half}_{\rm CIV}$} &
\colhead{loudness | R} & \colhead{DR7 T | DR7$+$ T | NIR T} \\
\colhead{} & \colhead{} & \colhead{(erg s$^{-1}$)}   & \colhead{(km s$^{-1}$)} & \colhead{($M_\odot$)} &
    \colhead{(\AA)}   & \colhead{(km s$^{-1}$ | km s$^{-1}$)} & 
    \colhead{flag | } & \colhead{}  \\
\colhead{(1)}   & \colhead{(2)}   & \colhead{(3)}   & \colhead{(4)}  & \colhead{(5)}  &
    \colhead{(6)}   & \colhead{(7)} & \colhead{(8)} & \colhead{(9)} }
\startdata
  J011521.20+152453.3 & 3.443 & 46.95 $\pm$ $<0.01$ &  6483 $\pm$  200 &  9.85 $\pm$ 0.03  &  27 $\pm$ 1    & 2433 $\pm$ 352 / 2678 $\pm$ 277 &-1 / -1    & 071205/ 071205 / 111022 \\   \\
  J014214.75+002324.2 & 3.379 & 47.01 $\pm$ $<0.01$ &  4730 $\pm$  227 &  9.61 $\pm$ 0.04  &  24 $\pm$ 1    &  962 $\pm$ 105 /  965 $\pm$ 100 & 0 / -1    & 000901/ 000901 / 111021\\   \\
  J015741.57-010629.6 & 3.572 & 46.99 $\pm$ $<0.01$ &  6489 $\pm$  149 &  9.87 $\pm$ 0.02  &  45 $\pm$ 1    & 1918 $\pm$  75 / 1590 $\pm$  61 & 0 / -1    & 001123/ 100910 / 111021 \\   \\
  J025021.76-075749.9 & 3.337 & 47.03 $\pm$ $<0.01$ &  6983 $\pm$  992 &  9.95 $\pm$ 0.12  &  18 $\pm$ 1    &  140 $\pm$ 776 / 1233 $\pm$ 233 & 0 / -1    & 001223/ 001223 / 111021 \\   \\
  J025905.63+001121.9 & 3.373 & 47.16 $\pm$ $<0.01$ &  5562 $\pm$  116 &  9.83 $\pm$ 0.02  &  55 $\pm$ 1    &  957 $\pm$  54 /  847 $\pm$  37 & 1 / 5.5   & 001223/ 101007 / 111021 \\   \\
  J030341.04-002321.9 & 3.233 & 47.16 $\pm$ $<0.01$ &  6383 $\pm$  386 &  9.95 $\pm$ 0.05  &  34 $\pm$ 1    &  511 $\pm$ 261 /  946 $\pm$ 217 & 0 / -1    & 000930/ 011023 / 111022 \\   \\
  J030449.85-000813.4 & 3.287 & 47.38 $\pm$ $<0.01$ &  3261 $\pm$  892 &  9.48 $\pm$ 0.19  &  21 $\pm$ 1    &  421 $\pm$ 137 /  506 $\pm$ 106 & 0 / -1    & 000930/ 000930 / 111020 \\   \\
  J075303.34+423130.8 & 3.590 & 47.22 $\pm$ $<0.01$ &  2903 $\pm$  731 &  9.29 $\pm$ 0.15  &  27 $\pm$ 1    &   22 $\pm$  93 /  117 $\pm$  42 & 1 / 2645.3& 000930/ 091213 / 111020\\   \\
  J080430.56+542041.1 & 3.759 & 47.10 $\pm$ $<0.01$ &  7041 $\pm$  253 & 10.00 $\pm$ 0.031 &  30 $\pm$ 1    & 1258 $\pm$ 155 / 1570 $\pm$ 102 & 0 / -1    & 050114/ 050114 / 111022\\   \\
  J080819.69+373047.3 & 3.480 & 46.85 $\pm$ $<0.01$ &  6714 $\pm$  341 &  9.82 $\pm$ 0.04  &  31 $\pm$ 1    & 1755 $\pm$ 120 / 1150 $\pm$ 149 & 0 / -1    & 011210/ 100312 / 111021\\   \\
  J080956.02+502000.9 & 3.281 & 47.03 $\pm$ $<0.01$ &  4016 $\pm$   73 &  9.48 $\pm$ 0.02  &  43 $\pm$ 1    &  147 $\pm$ 176 /  296 $\pm$ 163 & 0 / -1    & 040326/ 101117 / 111022\\   \\
  J081855.77+095848.0 & 3.700 & 47.09 $\pm$ $<0.01$ &  6860 $\pm$  479 &  9.97 $\pm$ 0.06  &  25 $\pm$ 1    & 1954 $\pm$ 282 / 1967 $\pm$ 168 & 0 / -1    & 070218/ 070218 / 111021/22\\   \\
  J090033.50+421547.0 & 3.290 & 47.49 $\pm$ $<0.01$ &  4888 $\pm$  522 &  9.89 $\pm$ 0.08  &  36 $\pm$ 1    & -360 $\pm$ 205 / -282 $\pm$ 196 & 1 / 2.1   & 020120/ 120226 / 120415\\   \\
  J094202.04+042244.5 & 3.276 & 47.37 $\pm$ $<0.01$ &  3356 $\pm$   51 &  9.50 $\pm$ 0.01  &  40 $\pm$ $<1$ &  372 $\pm$ 111 /  788 $\pm$ 110 & 0 / -1    & 011223/ 010411 / 120415\\   \\
  J115954.33+201921.1 & 3.426 & 47.40 $\pm$ $<0.01$ &  7525 $\pm$  310 & 10.21 $\pm$ 0.041 &  19 $\pm$ $<1$ & 1447 $\pm$ 533 / 1126 $\pm$  56 & 0 / -1    & 080109/ 080109 / 120416\\   \\
  J173352.23+540030.4 & 3.432 & 47.45 $\pm$ $<0.01$ &  5273 $\pm$  136 &  9.93 $\pm$ 0.02  &  16 $\pm$ 1    &  254 $\pm$ 712 / 1355 $\pm$ 703 & 1 / 14.0  & 000929/ 010331 / 120415\\   \\
  J213023.61+122252.0 & 3.272 & 47.10 $\pm$ $<0.01$ &  2474 $\pm$   77 &  9.09 $\pm$ 0.03  &  35 $\pm$ 1    &   16 $\pm$  31 /   46 $\pm$  36 &-1 / -1    & 020705/ 020705 / 111021\\   \\
  J224956.08+000218.0 & 3.311 & 46.90 $\pm$ $<0.01$ &  3160 $\pm$   42 &  9.20 $\pm$ 0.01  &  65 $\pm$ 1    &  226 $\pm$  35 /  490 $\pm$  25 & 0 / -1    & 021112/ 101002 / 111022\\   \\
  J230301.45-093930.7 & 3.492 & 47.27 $\pm$ $<0.01$ &  9453 $\pm$  453 & 10.34 $\pm$ 0.041 &  19 $\pm$ 1    & 1685 $\pm$ 540 / 1923 $\pm$ 392 & 0 / -1    & 011215/ 011215 / 111020\\   \\
\enddata
\tablecomments{
Col.(1) Name of the quasars.
Col.(2) Redshift measured from the H$\beta$ and the [OIII] doublets of the near-IR spectra.
Col.(3) Continuum luminosity at 1350 \AA.
Col.(4) FWHM of the CIV BC.
Col.(5) CIV line-based BH masses estimated using the calibration from \cite{Vestergaard_Peterson06}. The uncertainties quoted are only from statistical errors and not including the intrinsic uncertainties of the SE virial BH mass estimates as $\sim0.4$ dex \citep{Vestergaard02}.
Col.(6) REW of the CIV emission line.
Col.(7) CIV emission-line velocity shifts $\Delta V^{\rm peak}_{\rm CIV}$ and $\Delta V^{\rm half}_{\rm CIV}$ with respect to [OIII] $\lambda$5007.
Col.(8) FIRST radio flag: -1 = not in FIRST footprint; 0 = FIRST undetected; 1 = core dominant. Radio loudness R = $f_{\rm 6cm} / f_{2500}$. 
Col.(9) Date when the SDSS DR7 / SDSS DR7$+$ / Near-IR spectroscopy were taken, eg. 071110 referring to Nov. 10, 2007.}
\end{deluxetable*}

\begin{deluxetable}{cccccc}
\tabletypesize{\footnotesize}
\tablecaption{$\log M_{\rm BH}$ from the CIV emission line using different calibrations versus the reference H$\beta$-based $M_{\rm BH}$ estimates using the calibration in \cite{Vestergaard_Peterson06} \label{deltambh}}
\tablewidth{0pt}
\tablehead{
\colhead{$a$} & \colhead{$b$}  & \colhead{c} & \colhead{$\Delta \log M_{\rm BH}$}  & \colhead{$\sigma$} \\
\colhead{(1)}   & \colhead{(2)}   & \colhead{(3)}  & \colhead{(4)} & \colhead{(5)}   } 
\startdata
0.66            & 0.53 & 2.0 &  0.110   & 0.647 \\ \\
8.01 $\pm$ 0.26 & 0.53 &  0  &  0.025   & 0.360 \\ \\
2.26 $\pm$ 0.38 &  0   & 2.0 &  0.006   & 0.432 \\ \\
\enddata
\end{deluxetable}

\begin{deluxetable}{ccc}
\tabletypesize{\footnotesize}
\tablecaption{Correlations of the logarithm of the CIV-to-H$\beta$ BH mass ratios with Other Spectral Parameters \label{residual}}
\tablewidth{0pt}
\tablehead{
\colhead{Variable}   &  \colhead{$r$} & \colhead{$p$}  \\
\colhead{(1)}   & \colhead{(2)}   & \colhead{(3)}   }
\startdata
 $\log L_{1350}$                               &  0.31   & 0.19    \\
 $\log L_{5100}$                               &  0.04 & 0.86    \\
 $\log L_{1350} / L_{5100}$                    &  0.20   & 0.42    \\
 $\rm FWHM_{\rm CIV}$                          &  0.67   & $<0.01$ \\
 $\rm FWHM_{\rm H\beta}$                       & $-0.40$ & 0.09    \\
 $\rm FWHM_{\rm CIV} / FWHM_{\rm H\beta}$      &  0.98   & $<0.01$ \\ 
 CIV Blueshift $\Delta V_{\rm CIV}$            &  0.54   & 0.02    \\
 CIV Asymmetry AS$_{\rm CIV}$                  & $-0.62$ & $<0.01$    \\
\enddata
\tablecomments{Col.(2) The Spearman rank correlation coefficient $r$.
Col.(3) The significance of $r$ deviated from the null hypothesis.}
\end{deluxetable}

We thank the anonymous referee for his/her helpful comments that improved the paper. We thank the supports by the Ministry of Science and Technology of China under grant 2016YFA0400703, the NSFC grants No.11373008 and 11533001, and the National Key Basic Research Program of China 2014CB845700. This work made use of the facilities of the Center for High Performance Computing at Shanghai Astronomical Observatory, Chinese Academy of Sciences.
This research uses data obtained through the Telescope Access Program (TAP), which is funded by the National Astronomical Observatories, Chinese Academy of Sciences, and the Special Fund for Astronomy from the Ministry of Finance. Observations obtained with the Hale Telescope at Palomar Observatory were obtained as part of an agreement between the National Astronomical Observatories, Chinese Academy of Sciences, and the California Institute of Technology. The LBT is an international collabofn among institutions in the United States, Italy and Germany. LBT Corporation partners are: The University of Arizona on behalf of the Arizona university system; Istituto Nazionale di Astrofisica, Italy; LBT Beteiligungsgesellschaft, Germany, representing the Max-Planck Society, the Astrophysical Institute Potsdam, and Heidelberg University; The Ohio State University, and The Research Corporation, on behalf of The University of Notre Dame, University
of Minnesota and University of Virginia.
Funding for the SDSS and SDSS-II has been provided by the Alfred P. Sloan Foundation, the Participating Institutions, the National Science Foundation, the U.S. Department of Energy, the National Aeronautics and Space Administration, the Japanese Monbukagakusho, the Max Planck Society, and the Higher Education Funding Council for England. The SDSS Web Site is http://www.sdss.org/.
The SDSS is managed by the Astrophysical Research Consortium for the
Participating Institutions. The Participating Institutions are the
American Museum of Natural History, Astrophysical Institute Potsdam,
University of Basel, University of Cambridge, Case Western Reserve
University, University of Chicago, Drexel University, Fermilab, the
Institute for Advanced Study, the Japan Participation Group, Johns
Hopkins University, the Joint Institute for Nuclear Astrophysics, the
Kavli Institute for Particle Astrophysics and Cosmology, the Korean
Scientist Group, the Chinese Academy of Sciences (LAMOST), Los Alamos
National Laboratory, the Max-Planck-Institute for Astronomy (MPIA),
the Max- Planck-Institute for Astrophysics (MPA), New Mexico State
University, Ohio State University, University of Pittsburgh,
University of Portsmouth, Princeton University, the United States
Naval Observatory, and the University of Washington.


\begin{appendix}
In Model A, we fit 3 Gaussians for the BC of CIV, 1 Gaussian for the BC of HeII $\lambda1640$ and 
1 Gaussian for the BC of OIII$]$ $\lambda1663$.
In model B, we fit 2/1 Gaussians for the BC/NC of CIV, 1/1 Gaussian for the BC/NC of HeII and
1/1 Gaussian for the BC/NC of OIII]. The NC of the three lines are tied together and fitted with an 
upper FWHM limit as 1600 km s$^{-1}$.
We also compare them with Model C, where the only difference from Model A is that the number of 
Gaussians used to fit the CIV BC in Model C is 2 instead of 3.

Fig. \ref{fig:civspec_all_A} and Fig. \ref{fig:civspec_all_B} present the spectra fitting results 
of Model A and Model B in the wavelength range of 1500-1700 \AA\ for the 19 quasars, respectively. 
The reduced $\chi^{2}$ of the spectral fitting of the CIV emission line using Model A, B and C are 
listed in Table \ref{table:appendix-redchi2-compare}, with the median values as 1.07$\pm$0.25, 
1.06$\pm$0.27 and 1.16$\pm$0.34, respectively. Model A yields the smallest $\chi^2$ for 68.4\% quasars, 
though considering the scatter, the $\chi^2$ values of the three models are similar.

However, after visual inspections, we find that for most quasars the 3-Gaussian models
(Model A and B) for the CIV line reproduces the CIV line profile better than the 2-Gaussian Model (Model C). 
It is reasonable as there are more adjustable parameters to maximize the consistence between the model 
fitting and the observed data. For some quasars, such as $J011521.20+152453.3$ and 
$J014214.75+002324.2$, the fitting results from Model B show a weird NC due to the narrow spikes in 
the noisy regions of the spectra.

If the candidate model selected from the visual inspection (as listed in Col. (7) of Table \ref{table:appendix-redchi2-compare}) is different from the the best model according to the $\chi^2$ comparison
(as listed in Col. (6)), we adopt the candidate model from the visual inspection. If both Model A and Model B are listed in Col. (6) and Col. (7), we adopt the results from the Model A, which is supported by the Akaike information criterion (AIC) comparison, as listed in Col. (5).

To determine the best model for the data, we derive the AIC values of Model A
and Model B for each object. The optimal choice is the one with the smaller AIC value. The AIC is 
calculated as 
\begin{equation}
    \rm AIC = -2 \log L(\hat{\theta}) + 2 k,
\end{equation}
where $\theta$ is the set of model parameters, k is the number of the independently adjusted parameters in the 
candidate model, $L(\hat{\theta})$ is the likelihood of the candidate model giving the data when evaluated at the maximum likelihood estimate of $\theta$ \citep{Akaike_74}. The derived AIC values 
from Model A and Model B are listed in Table \ref{table:appendix-redchi2-compare}, showing that Model A is better to approximate the data than Model B.

In all, for 7 quasars ($J015741.57-010629.6$, $J025021.76-075749.9$, $J025905.63+001121.9$, 
$J030449.85-000813.4$, $J075303.34+423130.8$, $J080819.69+373047.3$ and $J090033.50+421547.0$), 
small fitting residuals around the CIV line are seen in the results from Model A but not shown in the results from 
Model B. Therefore, for these 7 quasars we use the results from Model B and for the other 12 quasars we adopt the 
results from Model A.

\begin{figure*}
\plotone{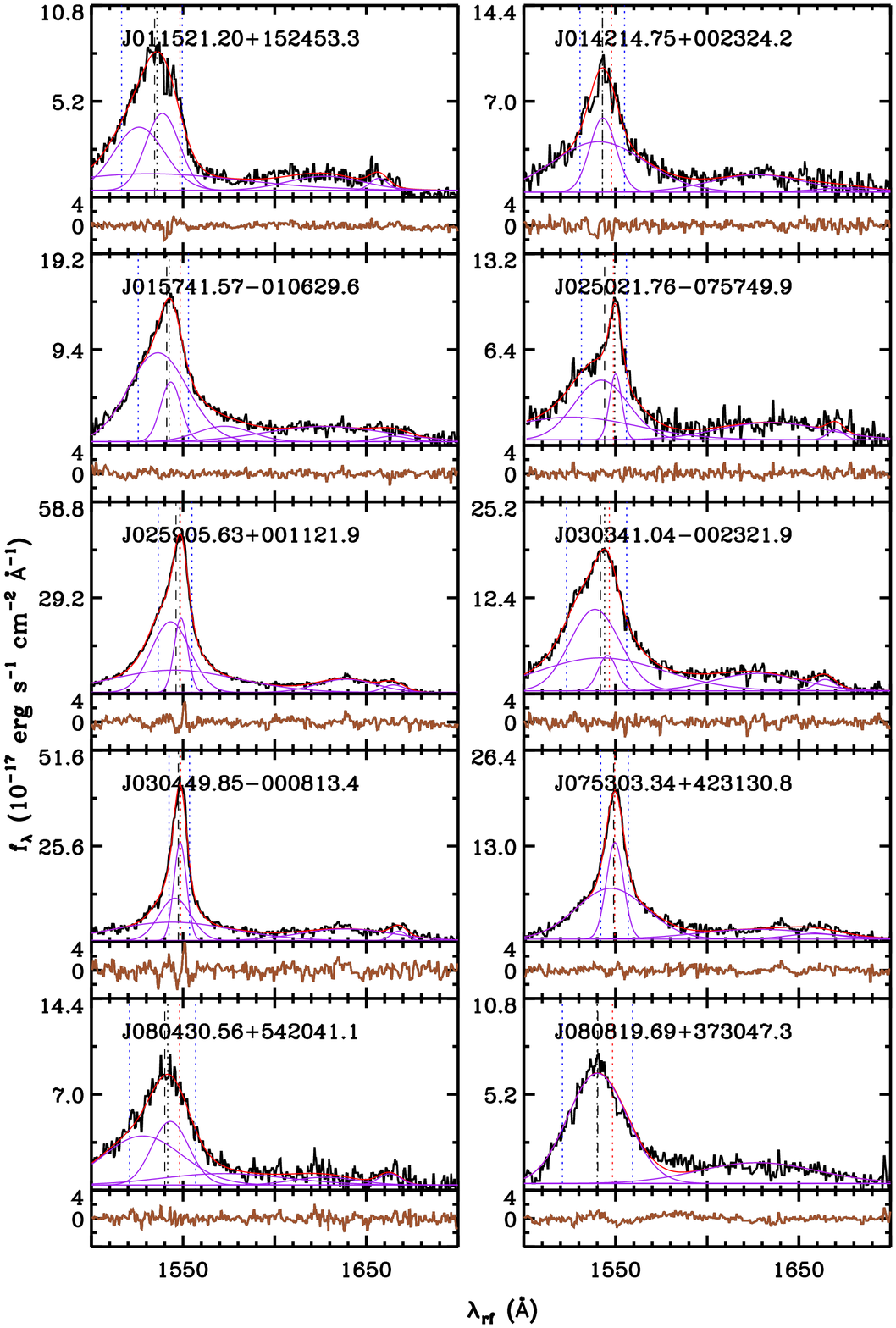}
\caption{Fitting results of the CIV line complex in the wavelength range of 1500-1700 \AA\  for the 19 quasars using Model A, where the spectrum in each panel is shown in black, the combined model fitting is shown in red, the individual Gaussian for the BC is shown in purple and the fitting residuals are shown in brown.  
The vertical blue dashed lines from left to right refer to the wavelengths of $\lambda_{\rm blue}$ 
and $\lambda_{\rm red}$. The vertical red dashed line refers to the wavelength of $\lambda_{\rm lab}$. 
The vertical black dashed and long dashed lines refer to the wavelength of $\lambda_{0}$ and $\lambda_{\rm half}$, respectively.}
\label{fig:civspec_all_A}
\end{figure*}
\renewcommand{\thefigure}{\arabic{figure}(Cont.)}
\addtocounter{figure}{-1}
\begin{figure*}
\plotone{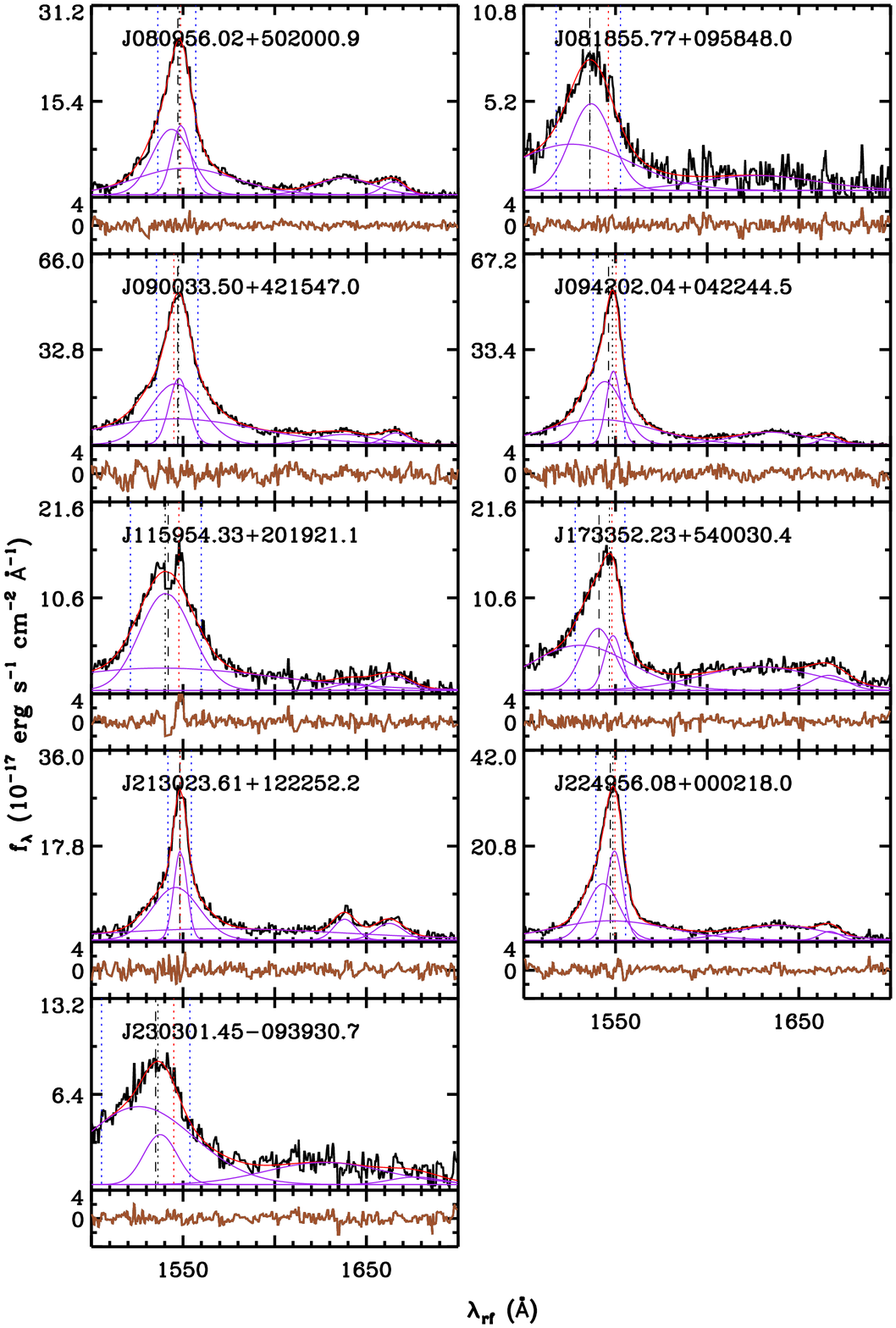}
\caption{(Continued)}
\end{figure*}
\renewcommand{\thefigure}{\arabic{figure}}

\begin{figure*}
\plotone{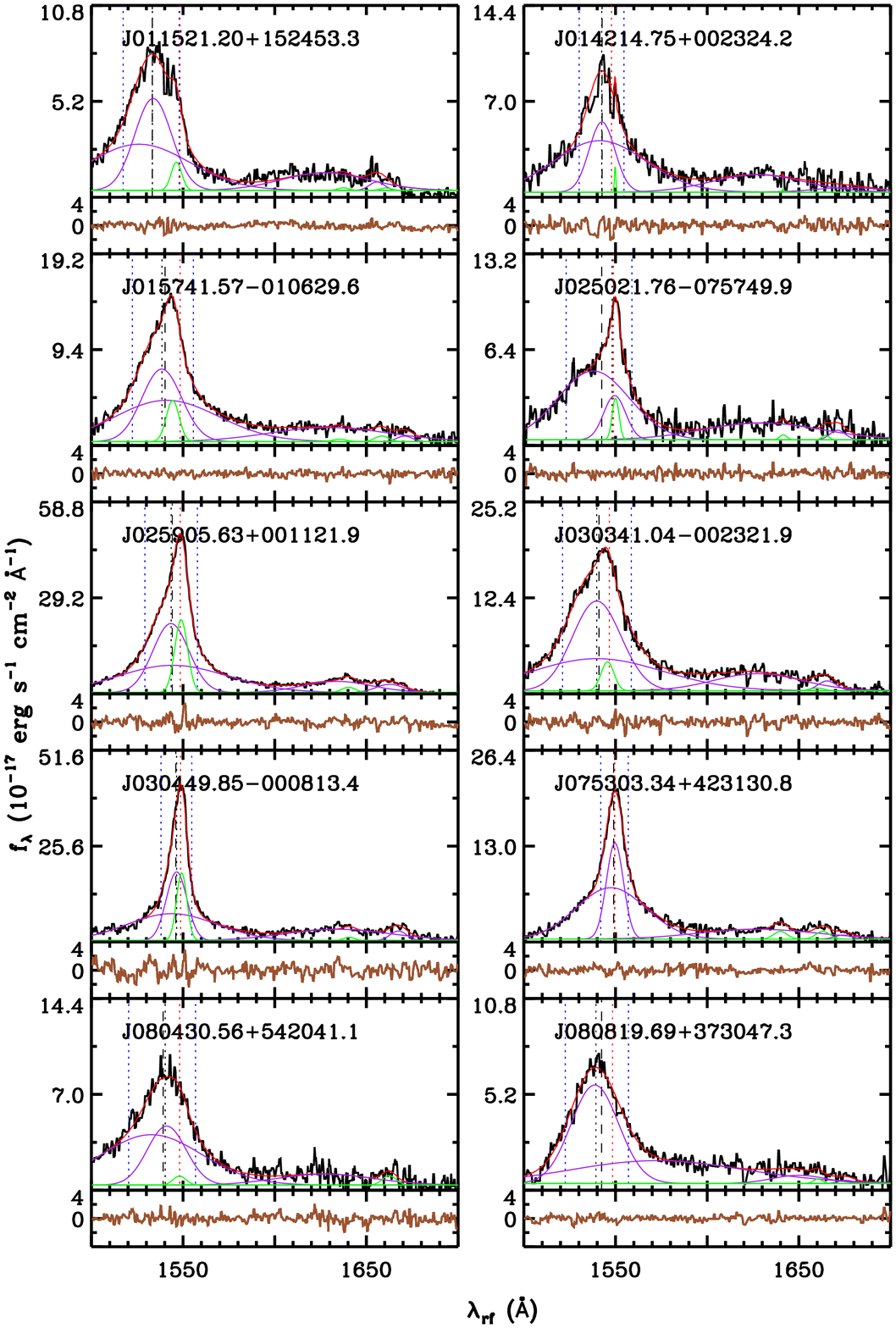}
\caption{Fitting results of the CIV line complex in the wavelength range of 1500-1700 \AA\  for the
19 quasars using Model B, where the spectrum in each panel is shown in black, the combined 
model fitting is shown in red, the individual Gaussian for the BC is shown in purple, the individual 
Gaussian for the NC is shown in green and the fitting residuals are shown in brown. 
The vertical blue dashed lines from left to right refer to the wavelengths of 
$\lambda_{\rm blue}$ and $\lambda_{\rm red}$. The vertical red dashed line refers to the wavelength of $\lambda_{\rm lab}$. 
The vertical black dashed and long dashed lines refer to the wavelength of $\lambda_{0}$ and $\lambda_{\rm half}$, respectively.}
\label{fig:civspec_all_B}
\end{figure*}
\renewcommand{\thefigure}{\arabic{figure}(Cont.)}
\addtocounter{figure}{-1}
\begin{figure*}
\plotone{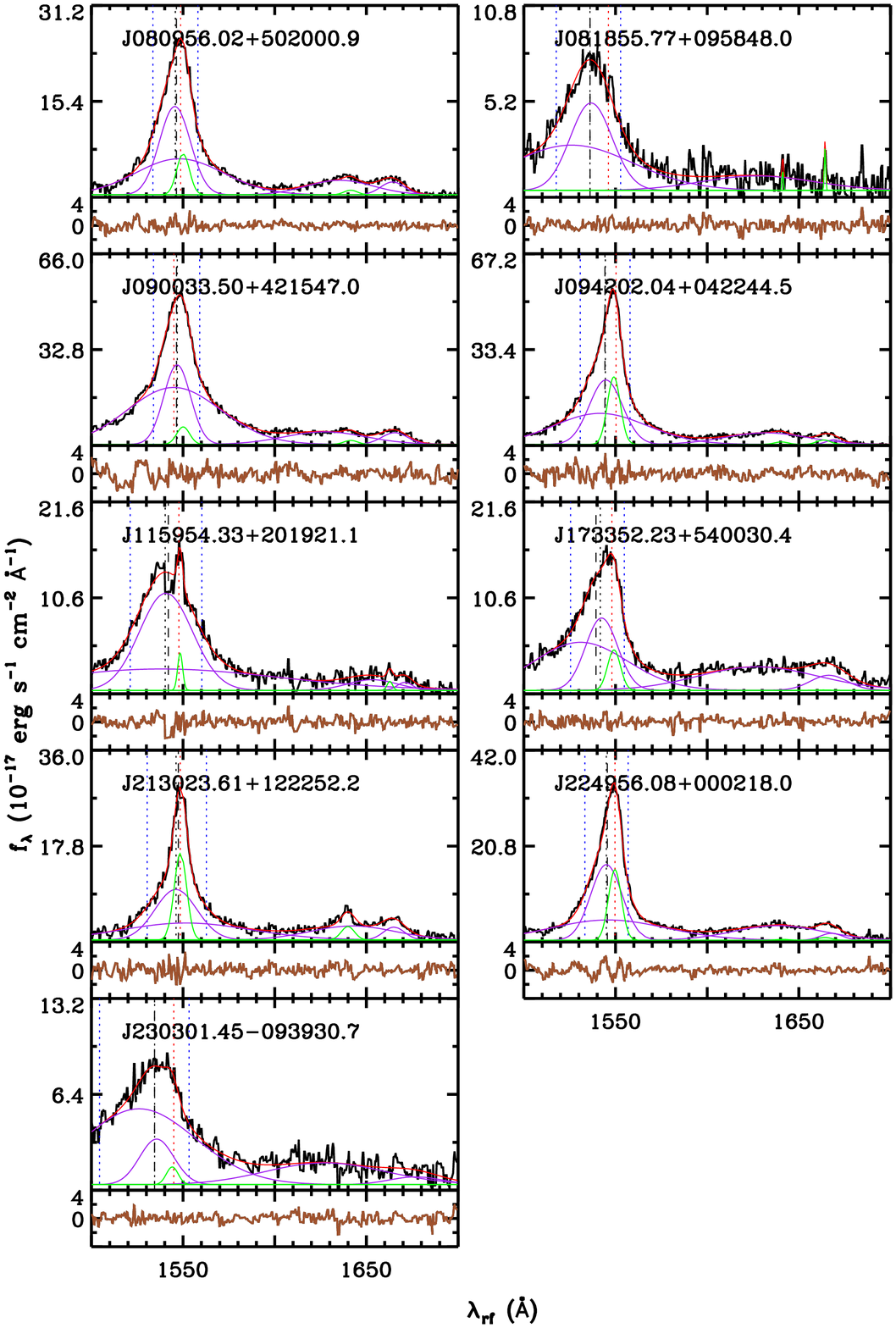}
\caption{(Continued)}
\end{figure*}
\renewcommand{\thefigure}{\arabic{figure}}

\begin{deluxetable}{ccccccccccccccccccc}
\tabletypesize{\footnotesize}
\tablecaption{The fitting details of the CIV emission line \label{table:appendix-redchi2-compare}}
\tablewidth{0pt}
\tablehead{
\colhead{Name (SDSS)} & \colhead{$\chi^{2}$} & \colhead{$\chi^{2}$} &
\colhead{$\chi^{2}$}   & \colhead{AIC$_{A}$-AIC$_{B}$} &
\colhead{$\chi^2$} & \colhead{Visual} & \colhead{Final} \\
\colhead{ } & \colhead{A} & \colhead{B} & \colhead{C} & 
\colhead{ } & \colhead{comparison} & \colhead{inspection} & \colhead{ }\\
\colhead{(1)} & \colhead{(2)} & \colhead{(3)} & \colhead{(4)} &
\colhead{(5)} & \colhead{(6)} & \colhead{(7)} & \colhead{(8)} \\} 
\startdata
 J011521.20+152453.3  & 1.29 &  1.30 &  1.34 & -4.001  &  A &  A  & A  \\ \\
 J014214.75+002324.2  & 1.21 &  1.20 &  1.24 & -4.004  &  B &  A  & A  \\ \\
 J015741.57-010629.6  & 0.96 &  0.88 &  1.08 & -3.998  &  B &  B  & B  \\ \\
 J025021.76-075749.9  & 0.96 &  0.96 &  0.98 & -3.999  &  AB&  B  & B  \\ \\
 J025905.63+001121.9  & 1.02 &  1.00 &  2.18 & -4.000  &  B &  AB & B  \\ \\
 J030341.04-002321.9  & 0.97 &  1.00 &  1.07 & -4.001  &  A &  A  & A  \\ \\
 J030449.85-000813.4  & 0.90 &  0.93 &  1.00 & -4.000  &  A &  B  & B  \\ \\
 J075303.34+423130.8  & 1.97 &  1.88 &  1.89 & -3.998  &  B &  AB & B  \\ \\
 J080430.56+542041.1  & 1.16 &  1.18 &  1.16 & -4.003  &  A &  A  & A  \\ \\
 J080819.69+373047.3  & 1.09 &  0.89 &  0.87 & -3.984  &  B &  B  & B  \\ \\
 J080956.02+502000.9  & 1.01 &  1.06 &  1.23 & -4.001  &  A &  A  & A  \\ \\
 J081855.77+095848.0  & 1.17 &  1.19 &  1.16 & -3.998  &  A &  A  & A  \\ \\
 J090033.50+421547.0  & 1.07 &  1.11 &  1.12 & -4.001  &  A &  B  & B  \\ \\
 J094202.04+042244.5  & 1.00 &  1.00 &  1.39 & -4.000  & AB &  AB & A  \\ \\
 J115954.33+201921.1  & 1.11 &  1.10 &  1.18 & -3.997  &  B &  A  & A  \\ \\
 J173352.23+540030.4  & 0.85 &  0.85 &  0.90 & -4.000  & AB &  AB & A  \\ \\
 J213023.61+122252.0  & 1.22 &  1.24 &  1.28 & -4.001  &  A &  AB & A  \\ \\
 J224956.08+000218.0  & 1.35 &  1.39 &  1.92 & -4.001  &  A &  A  & A  \\ \\
 J230301.45-093930.7  & 0.89 &  0.90 &  0.90 & -4.002  &  A &  A  & A  \\ \\
\enddata
\tablecomments{
Col.(1) Name of the quasars.
Col.(2) Reduced ${\chi}^{2}$ of the spectral fitting to the CIV line complex using Model A.
Col.(3) Reduced ${\chi}^{2}$ of the spectral fitting to the CIV line complex using Model B.
Col.(4) Reduced ${\chi}^{2}$ of the spectral fitting to the CIV line complex using Model C.
Col.(5) Difference of the AIC values between Model A and Model B.
Col.(6) The preferred model with the lower reduced $\chi^2$ between Model A and Model B.
Col.(7) The preferred model from the visual inspection between Model A and Model B.
Col.(8) The final adopted model from the combination of the results from Col.(5-7).
}
\end{deluxetable}
\end{appendix}

\begin{thebibliography}{}
\expandafter\ifx\csname natexlab\endcsname\relax\def\natexlab#1{#1}\fi

\bibitem[{{Akaike}(1974)}]{Akaike_74}
{Akaike}, H. 1974, IEEE Transactions on Automatic Control, 19, 716

\bibitem[{{Akritas} \& {Bershady}(1996)}]{Akritas_Bershady96}
{Akritas}, M.~G., \& {Bershady}, M.~A. 1996, \apj, 470, 706

\bibitem[{{Allen} {et~al.}(2013){Allen}, {Hewett}, {Richardson}, {Ferland}, \&
  {Baldwin}}]{Allen_etal_2013}
{Allen}, J.~T., {Hewett}, P.~C., {Richardson}, C.~T., {Ferland}, G.~J., \&
  {Baldwin}, J.~A. 2013, \mnras, 430, 3510

\bibitem[{{Assef} {et~al.}(2011){Assef}, {Denney}, {Kochanek}, {Peterson},
  {Koz{\l}owski}, {Ageorges}, {Barrows}, {Buschkamp}, {Dietrich}, {Falco},
  {Feiz}, {Gemperlein}, {Germeroth}, {Grier}, {Hofmann}, {Juette}, {Khan},
  {Kilic}, {Knierim}, {Laun}, {Lederer}, {Lehmitz}, {Lenzen}, {Mall}, {Madsen},
  {Mandel}, {Martini}, {Mathur}, {Mogren}, {Mueller}, {Naranjo}, {Pasquali},
  {Polsterer}, {Pogge}, {Quirrenbach}, {Seifert}, {Stern}, {Shappee}, {Storz},
  {Van Saders}, {Weiser}, \& {Zhang}}]{Assef_etal_11}
{Assef}, R.~J., {Denney}, K.~D., {Kochanek}, C.~S., {et~al.} 2011, \apj, 742,
  93

\bibitem[{{Bachev} {et~al.}(2004){Bachev}, {Marziani}, {Sulentic}, {Zamanov},
  {Calvani}, \& {Dultzin-Hacyan}}]{Bachev_etal_04}
{Bachev}, R., {Marziani}, P., {Sulentic}, J.~W., {et~al.} 2004, \apj, 617, 171

\bibitem[{{Baldwin}(1977)}]{baldwin_77}
{Baldwin}, J.~A. 1977, \apj, 214, 679

\bibitem[{{Barth} {et~al.}(2015){Barth}, {Bennert}, {Canalizo}, {Filippenko},
  {Gates}, {Greene}, {Li}, {Malkan}, {Pancoast}, {Sand }, {Stern}, {Treu},
  {Woo}, {Assef}, {Bae}, {Brewer}, {Cenko}, {Clubb}, {Cooper},
  {Diamond-Stanic}, {Hiner}, {H{\"o}nig}, {Hsiao}, {Kand rashoff}, {Lazarova},
  {Nierenberg}, {Rex}, {Silverman}, {Tollerud}, \& {Walsh}}]{Barth_etal_15}
{Barth}, A.~J., {Bennert}, V.~N., {Canalizo}, G., {et~al.} 2015, \apjs, 217, 26

\bibitem[{{Baskin} \& {Laor}(2005)}]{Baskin05}
{Baskin}, A., \& {Laor}, A. 2005, \mnras, 356, 1029

\bibitem[{{Becker} {et~al.}(2009){Becker}, {Rauch}, \& {Sargent}}]{Becker09}
{Becker}, G.~D., {Rauch}, M., \& {Sargent}, W.~L.~W. 2009, \apj, 698, 1010

\bibitem[{{Bentz} {et~al.}(2009){Bentz}, {Peterson}, {Netzer}, {Pogge}, \&
  {Vestergaard}}]{Bentz_etal_09}
{Bentz}, M.~C., {Peterson}, B.~M., {Netzer}, H., {Pogge}, R.~W., \&
  {Vestergaard}, M. 2009, \apj, 697, 160

\bibitem[{{Bentz} {et~al.}(2013){Bentz}, {Denney}, {Grier}, {Barth},
  {Peterson}, {Vestergaard}, {Bennert}, {Canalizo}, {De Rosa}, {Filippenko},
  {Gates}, {Greene}, {Li}, {Malkan}, {Pogge}, {Stern}, {Treu}, \&
  {Woo}}]{Bentz_etal_13}
{Bentz}, M.~C., {Denney}, K.~D., {Grier}, C.~J., {et~al.} 2013, \apj, 767, 149

\bibitem[{{Bian} {et~al.}(2010){Bian}, {Fan}, {Bechtold}, {McGreer}, {Just},
  {Sand}, {Green}, {Thompson}, {Peng}, {Seifert}, {Ageorges}, {Juette},
  {Knierim}, \& {Buschkamp}}]{Bian10}
{Bian}, F., {Fan}, X., {Bechtold}, J., {et~al.} 2010, \apj, 725, 1877

\bibitem[{{Boroson} \& {Green}(1992)}]{Boroson_Green92}
{Boroson}, T.~A., \& {Green}, R.~F. 1992, \apjs, 80, 109

\bibitem[{{Cardelli} {et~al.}(1989){Cardelli}, {Clayton}, \&
  {Mathis}}]{Cardelli89}
{Cardelli}, J.~A., {Clayton}, G.~C., \& {Mathis}, J.~S. 1989, \apj, 345, 245

\bibitem[{{Coatman} {et~al.}(2016){Coatman}, {Hewett}, {Banerji}, \&
  {Richards}}]{Coatman_etal_16}
{Coatman}, L., {Hewett}, P.~C., {Banerji}, M., \& {Richards}, G.~T. 2016,
  \mnras, 461, 647

\bibitem[{{Coatman} {et~al.}(2017){Coatman}, {Hewett}, {Banerji}, {Richards},
  {Hennawi}, \& {Prochaska}}]{Coatman_etal_17}
{Coatman}, L., {Hewett}, P.~C., {Banerji}, M., {et~al.} 2017, \mnras, 465, 2120

\bibitem[{{Coffey} {et~al.}(2019){Coffey}, {Salvato}, {Merloni}, {Boller},
  {Nandra}, {Dwelly}, {Comparat}, {Schulze}, {Del Moro}, \&
  {Schneider}}]{Coffey_etal_19}
{Coffey}, D., {Salvato}, M., {Merloni}, A., {et~al.} 2019, \aap, 625, A123

\bibitem[{{Cushing} {et~al.}(2004){Cushing}, {Vacca}, \& {Rayner}}]{Cushing04}
{Cushing}, M.~C., {Vacca}, W.~D., \& {Rayner}, J.~T. 2004, \pasp, 116, 362

\bibitem[{{Denney}(2012)}]{Denney12}
{Denney}, K.~D. 2012, \apj, 759, 44

\bibitem[{{Denney} {et~al.}(2010){Denney}, {Peterson}, {Pogge}, {Adair},
  {Atlee}, {Au-Yong}, {Bentz}, {Bird}, {Brokofsky}, {Chisholm}, {Comins},
  {Dietrich}, {Doroshenko}, {Eastman}, {Efimov}, {Ewald}, {Ferbey}, {Gaskell},
  {Hedrick}, {Jackson}, {Klimanov}, {Klimek}, {Kruse}, {Lad{\'e}route}, {Lamb},
  {Leighly}, {Minezaki}, {Nazarov}, {Onken}, {Petersen}, {Peterson},
  {Poindexter}, {Sakata}, {Schlesinger}, {Sergeev}, {Skolski}, {Stieglitz},
  {Tobin}, {Unterborn}, {Vestergaard}, {Watkins}, {Watson}, \&
  {Yoshii}}]{Denney_etal_10}
{Denney}, K.~D., {Peterson}, B.~M., {Pogge}, R.~W., {et~al.} 2010, \apj, 721,
  715

\bibitem[{{Du} {et~al.}(2014){Du}, {Hu}, {Lu}, {Wang}, {Qiu}, {Li}, {Bai},
  {Kaspi}, {Netzer}, {Wang}, \& {SEAMBH Collaboration}}]{Du_etal_14}
{Du}, P., {Hu}, C., {Lu}, K.-X., {et~al.} 2014, \apj, 782, 45

\bibitem[{{Elvis}(2000)}]{Elvis00}
{Elvis}, M. 2000, \apj, 545, 63

\bibitem[{{Feibelman}(1983)}]{Feibelman_83}
{Feibelman}, W.~A. 1983, \aap, 122, 335

\bibitem[{{Ferrarese} \& {Merritt}(2000)}]{Ferrarese00}
{Ferrarese}, L., \& {Merritt}, D. 2000, \apjl, 539, L9

\bibitem[{{Gaskell}(1982)}]{Gaskell_82}
{Gaskell}, C.~M. 1982, \apj, 263, 79

\bibitem[{{Ge} {et~al.}(2016){Ge}, {Bian}, {Jiang}, {Liu}, \&
  {Wang}}]{GeXue_etal_16}
{Ge}, X., {Bian}, W.-H., {Jiang}, X.-L., {Liu}, W.-S., \& {Wang}, X.-F. 2016,
  \mnras, 462, 966

\bibitem[{{Ge} {et~al.}(2019){Ge}, {Zhao}, {Bian}, \&
  {Frederick}}]{GeXue_etal_19}
{Ge}, X., {Zhao}, B.-X., {Bian}, W.-H., \& {Frederick}, G.~R. 2019, \aj, 157,
  148

\bibitem[{{Gebhardt} {et~al.}(2000){Gebhardt}, {Bender}, {Bower}, {Dressler},
  {Faber}, {Filippenko}, {Green}, {Grillmair}, {Ho}, {Kormendy}, {Lauer},
  {Magorrian}, {Pinkney}, {Richstone}, \& {Tremaine}}]{Gebhardt00}
{Gebhardt}, K., {Bender}, R., {Bower}, G., {et~al.} 2000, \apjl, 539, L13

\bibitem[{{Graham} {et~al.}(2011){Graham}, {Onken}, {Athanassoula}, \&
  {Combes}}]{Graham11}
{Graham}, A.~W., {Onken}, C.~A., {Athanassoula}, E., \& {Combes}, F. 2011,
  \mnras, 412, 2211

\bibitem[{{Greene} \& {Ho}(2005)}]{Greene05}
{Greene}, J.~E., \& {Ho}, L.~C. 2005, \apj, 630, 122

\bibitem[{{Greene} {et~al.}(2010){Greene}, {Peng}, \& {Ludwig}}]{Greene10}
{Greene}, J.~E., {Peng}, C.~Y., \& {Ludwig}, R.~R. 2010, \apj, 709, 937

\bibitem[{{Grier} {et~al.}(2017){Grier}, {Trump}, {Shen}, {Horne}, {Kinemuchi},
  {McGreer}, {Starkey}, {Brand t}, {Hall}, {Kochanek}, {Chen}, {Denney},
  {Greene}, {Ho}, {Homayouni}, {I-Hsiu Li}, {Pei}, {Peterson}, {Petitjean},
  {Schneider}, {Sun}, {AlSayyad}, {Bizyaev}, {Brinkmann}, {Brownstein},
  {Bundy}, {Dawson}, {Eftekharzadeh}, {Fernand ez-Trincado}, {Gao},
  {Hutchinson}, {Jia}, {Jiang}, {Oravetz}, {Pan}, {Paris}, {Ponder}, {Peters},
  {Rogerson}, {Simmons}, {Smith}, \& {Wang}}]{Grier_etal_17}
{Grier}, C.~J., {Trump}, J.~R., {Shen}, Y., {et~al.} 2017, \apj, 851, 21

\bibitem[{{Grier} {et~al.}(2019){Grier}, {Shen}, {Horne}, {Brandt}, {Trump},
  {Kinemuchi}, {Schneider}, {Homayouni}, {McGreer}, {Peterson}, {Bizyaev},
  {Chen}, {Dawson}, {Eftekharzadeh}, {Kneib}, {Nie}, {Oravetz}, {Oravetz},
  {Pan}, {Petitjean}, {Vivek}, \& {Zou}}]{Grier_etal_19}
{Grier}, C.~J., {Shen}, Y., {Horne}, K., {et~al.} 2019, arXiv e-prints,
  arXiv:1904.03199

\bibitem[{{G{\"u}ltekin} {et~al.}(2009){G{\"u}ltekin}, {Richstone}, {Gebhardt},
  {Lauer}, {Tremaine}, {Aller}, {Bender}, {Dressler}, {Faber}, {Filippenko},
  {Green}, {Ho}, {Kormendy}, {Magorrian}, {Pinkney}, \& {Siopis}}]{Gultekin09}
{G{\"u}ltekin}, K., {Richstone}, D.~O., {Gebhardt}, K., {et~al.} 2009, \apj,
  698, 198

\bibitem[{{Herter} {et~al.}(2008){Herter}, {Henderson}, {Wilson}, {Matthews},
  {Rahmer}, {Bonati}, {Muirhead}, {Adams}, {Lloyd}, {Skrutskie}, {Moon},
  {Parshley}, {Nelson}, {Martinache}, \& {Gull}}]{Terry08}
{Herter}, T.~L., {Henderson}, C.~P., {Wilson}, J.~C., {et~al.} 2008, in Society
  of Photo-Optical Instrumentation Engineers (SPIE) Conference Series, Vol.
  7014, Society of Photo-Optical Instrumentation Engineers (SPIE) Conference
  Series

\bibitem[{{Hill} {et~al.}(2012){Hill}, {Green}, {Ashby}, {Brynnel}, {Cushing},
  {Little}, {Slagle}, \& {Wagner}}]{Hill12}
{Hill}, J.~M., {Green}, R.~F., {Ashby}, D.~S., {et~al.} 2012, in Society of
  Photo-Optical Instrumentation Engineers (SPIE) Conference Series, Vol. 8444,
  Society of Photo-Optical Instrumentation Engineers (SPIE) Conference Series

\bibitem[{{Jiang} {et~al.}(2007){Jiang}, {Fan}, {Ivezi{\'c}}, {Richards},
  {Schneider}, {Strauss}, \& {Kelly}}]{Jiang2007}
{Jiang}, L., {Fan}, X., {Ivezi{\'c}}, {\v Z}., {et~al.} 2007, \apj, 656, 680

\bibitem[{{Kaspi} {et~al.}(2007){Kaspi}, {Brandt}, {Maoz}, {Netzer},
  {Schneider}, \& {Shemmer}}]{Kaspi_etal_07}
{Kaspi}, S., {Brandt}, W.~N., {Maoz}, D., {et~al.} 2007, \apj, 659, 997

\bibitem[{{Kaspi} {et~al.}(2005){Kaspi}, {Maoz}, {Netzer}, {Peterson},
  {Vestergaard}, \& {Jannuzi}}]{Kaspi_etal_05}
{Kaspi}, S., {Maoz}, D., {Netzer}, H., {et~al.} 2005, \apj, 629, 61

\bibitem[{{Kaspi} {et~al.}(2000){Kaspi}, {Smith}, {Netzer}, {Maoz}, {Jannuzi},
  \& {Giveon}}]{Kaspi_etal_00}
{Kaspi}, S., {Smith}, P.~S., {Netzer}, H., {et~al.} 2000, \apj, 533, 631

\bibitem[{{Kelly}(2007)}]{Kelly07}
{Kelly}, B.~C. 2007, \apj, 665, 1489

\bibitem[{{Konigl} \& {Kartje}(1994)}]{Konigl_Kartje_94}
{Konigl}, A., \& {Kartje}, J.~F. 1994, \apj, 434, 446

\bibitem[{{Kormendy} \& {Ho}(2013)}]{Kormendy13}
{Kormendy}, J., \& {Ho}, L.~C. 2013, \araa, 51, 511

\bibitem[{{Leighly}(2004)}]{Leighly04}
{Leighly}, K.~M. 2004, \apj, 611, 125

\bibitem[{{MacLeod} {et~al.}(2012){MacLeod}, {Ivezi{\'c}}, {Sesar}, {de Vries},
  {Kochanek}, {Kelly}, {Becker}, {Lupton}, {Hall}, {Richards}, {Anderson}, \&
  {Schneider}}]{MacLeod_etal_2012}
{MacLeod}, C.~L., {Ivezi{\'c}}, {\v{Z}}., {Sesar}, B., {et~al.} 2012, \apj,
  753, 106

\bibitem[{{Marziani} {et~al.}(2001){Marziani}, {Calvani}, \&
  {Braito}}]{Marziani_etal_01}
{Marziani}, P., {Calvani}, M., \& {Braito}, V. 2001, \memsai, 72, 41

\bibitem[{{Marziani} {et~al.}(1996){Marziani}, {Sulentic}, {Dultzin-Hacyan},
  {Calvani}, \& {Moles}}]{Marziani_etal_96}
{Marziani}, P., {Sulentic}, J.~W., {Dultzin-Hacyan}, D., {Calvani}, M., \&
  {Moles}, M. 1996, \apjs, 104, 37

\bibitem[{{Marziani} {et~al.}(2010){Marziani}, {Sulentic}, {Negrete},
  {Dultzin}, {Zamfir}, \& {Bachev}}]{Marziani_etal_10}
{Marziani}, P., {Sulentic}, J.~W., {Negrete}, C.~A., {et~al.} 2010, \mnras,
  409, 1033

\bibitem[{{Marziani} {et~al.}(2003){Marziani}, {Zamanov}, {Sulentic}, \&
  {Calvani}}]{Marziani_etal_03}
{Marziani}, P., {Zamanov}, R.~K., {Sulentic}, J.~W., \& {Calvani}, M. 2003,
  \mnras, 345, 1133

\bibitem[{{Marziani} {et~al.}(2019){Marziani}, {del Olmo},
  {Mart{\'\i}nez-Carballo}, {Mart{\'\i}nez-Aldama}, {Stirpe}, {Negrete},
  {Dultzin}, {D'Onofrio}, {Bon}, \& {Bon}}]{Marziani_etal_19}
{Marziani}, P., {del Olmo}, A., {Mart{\'\i}nez-Carballo}, M.~A., {et~al.} 2019,
  \aap, 627, A88

\bibitem[{{McGill} {et~al.}(2008){McGill}, {Woo}, {Treu}, \&
  {Malkan}}]{McGill08}
{McGill}, K.~L., {Woo}, J.-H., {Treu}, T., \& {Malkan}, M.~A. 2008, \apj, 673,
  703

\bibitem[{{McLure} \& {Jarvis}(2002)}]{McLure_Jarvis02}
{McLure}, R.~J., \& {Jarvis}, M.~J. 2002, \mnras, 337, 109

\bibitem[{{Mej{\'{\i}}a-Restrepo} {et~al.}(2018){Mej{\'{\i}}a-Restrepo},
  {Trakhtenbrot}, {Lira}, \& {Netzer}}]{Mejia_etal_18}
{Mej{\'{\i}}a-Restrepo}, J.~E., {Trakhtenbrot}, B., {Lira}, P., \& {Netzer}, H.
  2018, \mnras, 478, 1929

\bibitem[{{Murray} \& {Chiang}(1998)}]{Murray_Chiang_98}
{Murray}, N., \& {Chiang}, J. 1998, \apj, 494, 125

\bibitem[{{Murray} {et~al.}(1995){Murray}, {Chiang}, {Grossman}, \&
  {Voit}}]{Murray_etal_95}
{Murray}, N., {Chiang}, J., {Grossman}, S.~A., \& {Voit}, G.~M. 1995, \apj,
  451, 498

\bibitem[{{Nelson} {et~al.}(2004){Nelson}, {Green}, {Bower}, {Gebhardt}, \&
  {Weistrop}}]{Nelson04}
{Nelson}, C.~H., {Green}, R.~F., {Bower}, G., {Gebhardt}, K., \& {Weistrop}, D.
  2004, \apj, 615, 652

\bibitem[{{Onken} {et~al.}(2004){Onken}, {Ferrarese}, {Merritt}, {Peterson},
  {Pogge}, {Vestergaard}, \& {Wandel}}]{Onken04}
{Onken}, C.~A., {Ferrarese}, L., {Merritt}, D., {et~al.} 2004, \apj, 615, 645

\bibitem[{{Park} {et~al.}(2017){Park}, {Barth}, {Woo}, {Malkan}, {Treu},
  {Bennert}, {Assef}, \& {Pancoast}}]{Park_etal_17}
{Park}, D., {Barth}, A.~J., {Woo}, J.-H., {et~al.} 2017, \apj, 839, 93

\bibitem[{{Park} {et~al.}(2013){Park}, {Woo}, {Denney}, \&
  {Shin}}]{Park_etal_13}
{Park}, D., {Woo}, J.-H., {Denney}, K.~D., \& {Shin}, J. 2013, \apj, 770, 87

\bibitem[{{Peterson}(1993)}]{Peterson93}
{Peterson}, B.~M. 1993, \pasp, 105, 247

\bibitem[{{Peterson} {et~al.}(2004){Peterson}, {Ferrarese}, {Gilbert}, {Kaspi},
  {Malkan}, {Maoz}, {Merritt}, {Netzer}, {Onken}, {Pogge}, {Vestergaard}, \&
  {Wandel}}]{Peterson04}
{Peterson}, B.~M., {Ferrarese}, L., {Gilbert}, K.~M., {et~al.} 2004, \apj, 613,
  682

\bibitem[{{Proga} {et~al.}(2000){Proga}, {Stone}, \& {Kallman}}]{Proga_etal_00}
{Proga}, D., {Stone}, J.~M., \& {Kallman}, T.~R. 2000, \apj, 543, 686

\bibitem[{{Richards} {et~al.}(2002){Richards}, {Vanden Berk}, {Reichard},
  {Hall}, {Schneider}, {SubbaRao}, {Thakar}, \& {York}}]{Richards02}
{Richards}, G.~T., {Vanden Berk}, D.~E., {Reichard}, T.~A., {et~al.} 2002, \aj,
  124, 1

\bibitem[{{Richards} {et~al.}(2006){Richards}, {Lacy}, {Storrie-Lombardi},
  {Hall}, {Gallagher}, {Hines}, {Fan}, {Papovich}, {Vanden Berk}, {Trammell},
  {Schneider}, {Vestergaard}, {York}, {Jester}, {Anderson}, {Budav{\'a}ri}, \&
  {Szalay}}]{Richards_etal_06}
{Richards}, G.~T., {Lacy}, M., {Storrie-Lombardi}, L.~J., {et~al.} 2006, \apjs,
  166, 470

\bibitem[{{Richards} {et~al.}(2011){Richards}, {Kruczek}, {Gallagher}, {Hall},
  {Hewett}, {Leighly}, {Deo}, {Kratzer}, \& {Shen}}]{Richards_etal_11}
{Richards}, G.~T., {Kruczek}, N.~E., {Gallagher}, S.~C., {et~al.} 2011, \aj,
  141, 167

\bibitem[{{Runnoe} {et~al.}(2013){Runnoe}, {Brotherton}, {Shang}, \&
  {DiPompeo}}]{Runnoe13}
{Runnoe}, J.~C., {Brotherton}, M.~S., {Shang}, Z., \& {DiPompeo}, M.~A. 2013,
  \mnras, arXiv:1306.3521

\bibitem[{{Schlegel} {et~al.}(1998){Schlegel}, {Finkbeiner}, \&
  {Davis}}]{Schlegel98}
{Schlegel}, D.~J., {Finkbeiner}, D.~P., \& {Davis}, M. 1998, \apj, 500, 525

\bibitem[{{Schneider} {et~al.}(2010){Schneider}, {Richards}, {Hall}, {Strauss},
  {Anderson}, {Boroson}, {Ross}, {Shen}, {Brandt}, {Fan}, {Inada}, {Jester},
  {Knapp}, {Krawczyk}, {Thakar}, {Vanden Berk}, {Voges}, {Yanny}, {York},
  {Bahcall}, {Bizyaev}, {Blanton}, {Brewington}, {Brinkmann}, {Eisenstein},
  {Frieman}, {Fukugita}, {Gray}, {Gunn}, {Hibon}, {Ivezi{\'c}}, {Kent}, {Kron},
  {Lee}, {Lupton}, {Malanushenko}, {Malanushenko}, {Oravetz}, {Pan}, {Pier},
  {Price}, {Saxe}, {Schlegel}, {Simmons}, {Snedden}, {SubbaRao}, {Szalay}, \&
  {Weinberg}}]{Schneider10}
{Schneider}, D.~P., {Richards}, G.~T., {Hall}, P.~B., {et~al.} 2010, \aj, 139,
  2360

\bibitem[{{Schulze} {et~al.}(2018){Schulze}, {Silverman}, {Kashino}, {Akiyama},
  {Schramm}, {Sanders}, {Kartaltepe}, {Daddi}, {Rodighiero}, {Renzini},
  {Arimoto}, {Nagao}, {Puglisi}, {Trakhtenbrot}, {Civano}, \&
  {Suh}}]{Schulze18}
{Schulze}, A., {Silverman}, J.~D., {Kashino}, D., {et~al.} 2018, \apjs, 239, 22

\bibitem[{{Shemmer} \& {Lieber}(2015)}]{Shemmer_Lieber_15}
{Shemmer}, O., \& {Lieber}, S. 2015, \apj, 805, 124

\bibitem[{{Shen}(2013)}]{Shen13}
{Shen}, Y. 2013, Bulletin of the Astronomical Society of India, 41, 61

\bibitem[{{Shen} {et~al.}(2008){Shen}, {Greene}, {Strauss}, {Richards}, \&
  {Schneider}}]{Shen08}
{Shen}, Y., {Greene}, J.~E., {Strauss}, M.~A., {Richards}, G.~T., \&
  {Schneider}, D.~P. 2008, \apj, 680, 169

\bibitem[{{Shen} \& {Ho}(2014)}]{Shen_Ho14}
{Shen}, Y., \& {Ho}, L.~C. 2014, \nat, 513, 210

\bibitem[{{Shen} \& {Liu}(2012)}]{Shen_Liu12}
{Shen}, Y., \& {Liu}, X. 2012, \apj, 753, 125

\bibitem[{{Shen} {et~al.}(2011){Shen}, {Richards}, {Strauss}, {Hall},
  {Schneider}, {Snedden}, {Bizyaev}, {Brewington}, {Malanushenko},
  {Malanushenko}, {Oravetz}, {Pan}, \& {Simmons}}]{Shen11}
{Shen}, Y., {Richards}, G.~T., {Strauss}, M.~A., {et~al.} 2011, \apjs, 194, 45

\bibitem[{{Shen} {et~al.}(2015){Shen}, {Brandt}, {Dawson}, {Hall}, {McGreer},
  {Anderson}, {Chen}, {Denney}, {Eftekharzadeh}, {Fan}, {Gao}, {Green},
  {Greene}, {Ho}, {Horne}, {Jiang}, {Kelly}, {Kinemuchi}, {Kochanek},
  {P{\^a}ris}, {Peters}, {Peterson}, {Petitjean}, {Ponder}, {Richards},
  {Schneider}, {Seth}, {Smith}, {Strauss}, {Tao}, {Trump}, {Wood-Vasey}, {Zu},
  {Eisenstein}, {Pan}, {Bizyaev}, {Malanushenko}, {Malanushenko}, \&
  {Oravetz}}]{Shen_etal_15}
{Shen}, Y., {Brandt}, W.~N., {Dawson}, K.~S., {et~al.} 2015, \apjs, 216, 4

\bibitem[{{Shen} {et~al.}(2019){Shen}, {Grier}, {Horne}, {Brandt}, {Trump},
  {Hall}, {Kinemuchi}, {Starkey}, {Schneider}, {Ho}, {Homayouni}, {I-Hsiu Li},
  {McGreer}, {Peterson}, {Bizyaev}, {Chen}, {Dawson}, {Eftekharzadeh}, {Green},
  {Guo}, {Jia}, {Jiang}, {Kneib}, {Li}, {Li}, {Nie}, {Oravetz}, {Oravetz},
  {Pan}, {Petitjean}, {Ponder}, {Rogerson}, {Vivek}, {Zhang}, \&
  {Zou}}]{Shen_etal_19}
{Shen}, Y., {Grier}, C.~J., {Horne}, K., {et~al.} 2019, \apjl, 883, L14

\bibitem[{{Sulentic} {et~al.}(2007){Sulentic}, {Bachev}, {Marziani}, {Negrete},
  \& {Dultzin}}]{Sulentic_etal_07}
{Sulentic}, J.~W., {Bachev}, R., {Marziani}, P., {Negrete}, C.~A., \&
  {Dultzin}, D. 2007, \apj, 666, 757

\bibitem[{{Sulentic} {et~al.}(2000){Sulentic}, {Marziani}, \&
  {Dultzin-Hacyan}}]{Sulentic_etal_00}
{Sulentic}, J.~W., {Marziani}, P., \& {Dultzin-Hacyan}, D. 2000, \araa, 38, 521

\bibitem[{{Sulentic} {et~al.}(2017){Sulentic}, {del Olmo}, {Marziani},
  {Mart{\'{\i}}nez-Carballo}, {D'Onofrio}, {Dultzin}, {Perea},
  {Mart{\'{\i}}nez-Aldama}, {Negrete}, {Stirpe}, \&
  {Zamfir}}]{Sulentic_etal_17}
{Sulentic}, J.~W., {del Olmo}, A., {Marziani}, P., {et~al.} 2017, \aap, 608,
  A122

\bibitem[{{Sun} {et~al.}(2018){Sun}, {Xue}, {Richards}, {Trump}, {Shen},
  {Brandt}, \& {Schneider}}]{Sun_etal_18}
{Sun}, M., {Xue}, Y., {Richards}, G.~T., {et~al.} 2018, \apj, 854, 128

\bibitem[{{Trakhtenbrot} \& {Netzer}(2012)}]{Trakhtenbrot12}
{Trakhtenbrot}, B., \& {Netzer}, H. 2012, \mnras, 427, 3081

\bibitem[{{Tremaine} {et~al.}(2002){Tremaine}, {Gebhardt}, {Bender}, {Bower},
  {Dressler}, {Faber}, {Filippenko}, {Green}, {Grillmair}, {Ho}, {Kormendy},
  {Lauer}, {Magorrian}, {Pinkney}, \& {Richstone}}]{Tremaine02}
{Tremaine}, S., {Gebhardt}, K., {Bender}, R., {et~al.} 2002, \apj, 574, 740

\bibitem[{{Vanden Berk} {et~al.}(2001){Vanden Berk}, {Richards}, {Bauer},
  {Strauss}, {Schneider}, {Heckman}, {York}, {Hall}, {Fan}, {Knapp},
  {Anderson}, {Annis}, {Bahcall}, {Bernardi}, {Briggs}, {Brinkmann}, {Brunner},
  {Burles}, {Carey}, {Castander}, {Connolly}, {Crocker}, {Csabai}, {Doi},
  {Finkbeiner}, {Friedman}, {Frieman}, {Fukugita}, {Gunn}, {Hennessy},
  {Ivezi{\'c}}, {Kent}, {Kunszt}, {Lamb}, {Leger}, {Long}, {Loveday}, {Lupton},
  {Meiksin}, {Merelli}, {Munn}, {Newberg}, {Newcomb}, {Nichol}, {Owen}, {Pier},
  {Pope}, {Rockosi}, {Schlegel}, {Siegmund}, {Smee}, {Snir}, {Stoughton},
  {Stubbs}, {SubbaRao}, {Szalay}, {Szokoly}, {Tremonti}, {Uomoto}, {Waddell},
  {Yanny}, \& {Zheng}}]{Berk_etal_01}
{Vanden Berk}, D.~E., {Richards}, G.~T., {Bauer}, A., {et~al.} 2001, \aj, 122,
  549

\bibitem[{{Vestergaard}(2002)}]{Vestergaard02}
{Vestergaard}, M. 2002, \apj, 571, 733

\bibitem[{{Vestergaard} \& {Osmer}(2009)}]{Vestergaard_Osmer09}
{Vestergaard}, M., \& {Osmer}, P.~S. 2009, \apj, 699, 800

\bibitem[{{Vestergaard} \& {Peterson}(2006)}]{Vestergaard_Peterson06}
{Vestergaard}, M., \& {Peterson}, B.~M. 2006, \apj, 641, 689

\bibitem[{{Vietri} {et~al.}(2018){Vietri}, {Piconcelli}, {Bischetti}, {Duras},
  {Martocchia}, {Bongiorno}, {Marconi}, {Zappacosta}, {Bisogni}, {Bruni},
  {Brusa}, {Comastri}, {Cresci}, {Feruglio}, {Giallongo}, {La Franca},
  {Mainieri}, {Mannucci}, {Ricci}, {Sani}, {Testa}, {Tombesi}, {Vignali}, \&
  {Fiore}}]{Vietri_etal_18}
{Vietri}, G., {Piconcelli}, E., {Bischetti}, M., {et~al.} 2018, \aap, 617, A81

\bibitem[{{Wilhite} {et~al.}(2007){Wilhite}, {Brunner}, {Schneider}, \& {Vanden
  Berk}}]{Wilhite_etal_2007}
{Wilhite}, B.~C., {Brunner}, R.~J., {Schneider}, D.~P., \& {Vanden Berk}, D.~E.
  2007, \apj, 669, 791

\bibitem[{{Zuo} {et~al.}(2015){Zuo}, {Wu}, {Fan}, {Green}, {Wang}, \&
  {Bian}}]{Zuo_etal_15}
{Zuo}, W., {Wu}, X.-B., {Fan}, X., {et~al.} 2015, \apj, 799, 189

\bibitem[{{Zuo} {et~al.}(2012){Zuo}, {Wu}, {Liu}, \& {Jiao}}]{Zuo_etal_2012}
{Zuo}, W., {Wu}, X.-B., {Liu}, Y.-Q., \& {Jiao}, C.-L. 2012, \apj, 758, 104

\end{thebibliography}
\end{document}